\documentclass{article}
\usepackage{amsmath,graphicx}
\usepackage[a4paper]{geometry}
\usepackage{xcolor}
\usepackage{amsfonts,authblk}
\usepackage{hyperref}
\usepackage{subfigure}
\usepackage{soul}

\author[1]{Miko\l aj Lasota\thanks{\href{mailto:miklas@umk.pl}{Corresponding author: miklas@umk.pl}}}
\author[2]{Olena Kovalenko}
\author[2]{Vladyslav C. Usenko}
\affil[1]{Faculty of Physics, Astronomy and Informatics, Nicolaus Copernicus University, Grudziadzka 5, 87-100 Toru\'{n}, Poland}
\affil[2]{Department of Optics, Palack\'{y} University, 17.\,listopadu 1192/12, 77146 Olomouc, Czech Republic}

\begin{document}

\title{Robustness of entanglement-based discrete- and continuous-variable quantum key distribution against channel noise}

\maketitle

\begin{abstract}
Discrete-variable (DV) and continuous-variable (CV) schemes constitute the two major families of quantum key distribution (QKD) protocols. Unfortunately, since the setup elements required by these schemes are quite different, making a fair comparison of their potential performance in particular applications is often troublesome, limiting the experimenters' capability to choose an optimal solution. In this work we perform a general comparison of the major entanglement-based DV and CV QKD protocols in terms of their resistance to the channel noise, with the otherwise perfect setup, showing the definite superiority of the DV family. We analytically derive fundamental bounds on the tolerable channel noise and attenuation for entanglement-based CV QKD protocols. We also investigate the influence of DV QKD setup imperfections on the obtained results in order to determine benchmarks for the parameters of realistic photon sources and detectors, allowing the realistic DV protocols to outperform even the ideal CV QKD analogs. Our results indicate the realistic advantage of DV entanglement-based schemes over their CV counterparts and suggests the practical efforts for maximizing this advantage.
 
\end{abstract}

\section{Introduction}
Quantum key distribution (QKD) \cite{Pirandola2020} is one of the most advanced quantum technologies, having its goal in development of protocols for provably secure distribution of random secret keys, that can be used for classical symmetric cryptosystems. It was first suggested on the basis of single-photon states \cite{Bennett84} and tested using weak coherent pulses \cite{Bennett1992} in the so-called prepare-and-measure (P\&M) design, where one of the trusted parties prepares the signal states and sends them through an untrusted channel to the remote trusted party to perform the measurements on the signal. QKD was later extended to the entanglement-based configuration \cite{Ekert1991} (also often referred to as EPR-based, which links to the famous Eintstein-Podolsky-Rosen paradox of the entangled states \cite{Einstein1935}). While in the basic case EPR-based QKD schemes can be seen as equivalent to the P\&M ones, they can also offer higher robustness against channel loss compared to practical P\&M realizations \cite{Ma2007}, device-independent security based on the Bell inequality violation \cite{Vazirani2014}, possibility for utilization in quantum repeater schemes for extending secure distance using entanglement swapping \cite{Sangouard2011}, as well as scalability which allows performing QKD between multiple users \cite{Xue2017,Wengerowsky2018}. Nowadays QKD schemes based on twin-field \cite{Lucamarini2018} or measurement-device-independent (MDI) \cite{Lo2012,Zeng2022,Xie2022} protocols are capable of distributing secure cryptographic keys on the distances of several hundred \cite{Zhou2023} or even one thousand \cite{Liu2023} kilometers, at least in laboratory conditions.
	
As an alternative to discrete-variable (DV) QKD protocols employing direct photodetection, mentioned above, the continuous-variable (CV) QKD \cite{Braunstein2005} was suggested based on efficient and fast homodyne detection \cite{Zhang2018} of generally multiphoton quantum states. Successfully realized in P\&M scenario \cite{Jouguet2013}, it was also tested in the EPR-based scheme \cite{Madsen2012}. While CV QKD protocols aim primarily at high-speed and efficient key generation, their performance is limited by losses and noise in the channel, as well as imperfect error correction and flawed practical devices \cite{Jouguet2012}. Previously the DV and CV P\&M protocols were analyzed and compared in terms of their robustness to excess noise in untrusted quantum channels, revealing certain advantage of CV QKD in middle ranges, overwhelmed by DV QKD in long-distance channels \cite{Lasota2017}. The analysis is relevant, as untrusted channel noise can be present, \emph{e.g.} in practical case of co-existence between quantum and strong classical signals, where crosstalk effects can lead to noise leakage to the quantum channel \cite{Eraerds2010}. Considering the higher level of security and scalability offered by the EPR-based QKD realizations, mentioned above, an analogous comparison done in the entanglement-based regime would be even more useful for determining the optimal QKD approaches in practical situations. However, according to our knowledge such study has not been performed yet. 
	
In this paper we fill this gap by analyzing and comparing the performance of EPR-based DV and CV protocols in noisy quantum channels, with otherwise perfect setups. We study standard EPR-based protocols with entanglement source placed in the middle of the channel (typical networking configuration usually considered for DV QKD and also suggested for CV QKD \cite{Weedbrook2013}) as well as the MDI protocols, in which the Bell-type detection is given to an eavesdropper as the part of the channel \cite{Lo2012,Braunstein2012}. In both scenarios we analytically derive the fundamental bounds on the channel noise and attenuation for the CV QKD protocols, the latter corresponding respectively to approximately 10 and 7 kilometers of standard telecom fiber in each arm of the entangled state. Furthermore, we address the robustness of the fully device-independent (DI) DV QKD, but make no comparison as no CV counterpart of the protocol is known. Our results show the general superiority of DV EPR-based standard and MDI QKD protocols over the CV QKD ones in terms of robustness to the channel noise, caused mainly by the fact that EPR-based CV QKD cannot tolerate more than -3 dB loss in either of the channels \cite{Usenko2016}. Basing on this outcome, we also consider various imperfections of realistic photon sources and detectors used by the DV schemes in order to determine benchmarks for their parameters that guarantee superiority of practical EPR DV QKD schemes over even the ideal CV analogs.
	
Since we are focused on comparison of the ultimate performance of perfectly implemented QKD protocols for setting the bounds achievable by both CV and DV families of EPR-based protocols, we consider asymptotic security in the assumption of infinitely many signals shared between the trusted parties. However, practical implementations always deal with the finite-size regime, when the data ensembles are limited, and the protocol optimizations should be performed in both DV \cite{Cai2009} and CV \cite{Leverrier2010} regimes to achieve maximum performance of the practical QKD protocols, compatible with the predefined failure probabilities. Nevertheless, one can expect that in the low-loss / strong-noise scenario, that is our primary focus in this work, the legitimate parties should typically be able to generate very long raw keys relatively fast and therefore analysing the finite-size effects would not be critical in this situation.
	
The paper is composed as follows. In Sec.\ref{Sec:DVtheory} we introduce the two considered DV QKD setup configurations, elaborate on the assumptions on the individual setup elements and perform the security analysis of the DV QKD protocols realized in these configurations. Next, in Sec.\ref{Sec:CVtheory}, we introduce the CV QKD analogs of the aforementioned setups and analyze the security of the QKD protocols realized with their use. The comparison between the performance of different protocols utilizing ideal light sources and detectors is performed in Sec.\ref{Sec:Results}. Also the benchmarks for the parameters of realistic photon-pair sources and photon detectors guaranteeing the superiority of the DV protocols over their CV competitors are investigated there. Finally, Sec.\ref{Sec:Summary} provides the summary of our work.
	
\section{Security of EPR-based DV QKD}
\label{Sec:DVtheory}
	
Security of DV QKD is assessed as the positivity of the lower bound on the secure key rate per channel use, which can be written as \cite{Scarani09}
\begin{equation}
    K^{DV}=p_\mathrm{exp}\cdot\Delta I,
\end{equation}
where $p_\mathrm{exp}$ denotes the expected probability for accepting a given attempt by the legitimate parties and 
\begin{equation}
    \Delta I=\max\left[0,I_{AB}-\min\{I_{EA},I_{EB}\}\right]
\label{eq:KDV}
\end{equation}
is called the secret fraction \cite{Csiszar78,Devetak05}. In the above formula $I_{AB}$ is the mutual information between Alice and Bob, and $I_{EA}$ ($I_{EB}$) represents the amount of information Eve can gain on Alice’s (Bob’s) raw key bit upon an eavesdropping attack. Specific conditions for acceptance of a given key generation attempt by the trusted parties depend on the utilized QKD scheme. Below we analyze the security of two QKD setup configurations with a central station located in between Alice and Bob, which correspond to two different types of protocols, namely the standard ones, when the central station is equipped with a source, and the measurement-device independent (MDI) protocols, where the central station is equipped with a set of detectors instead.
	
\subsection{Source of entanglement in the central station}
\label{Sec:SourceMiddle}

\begin{figure}[tbh]
    \centering
    \includegraphics[width=0.8\textwidth]{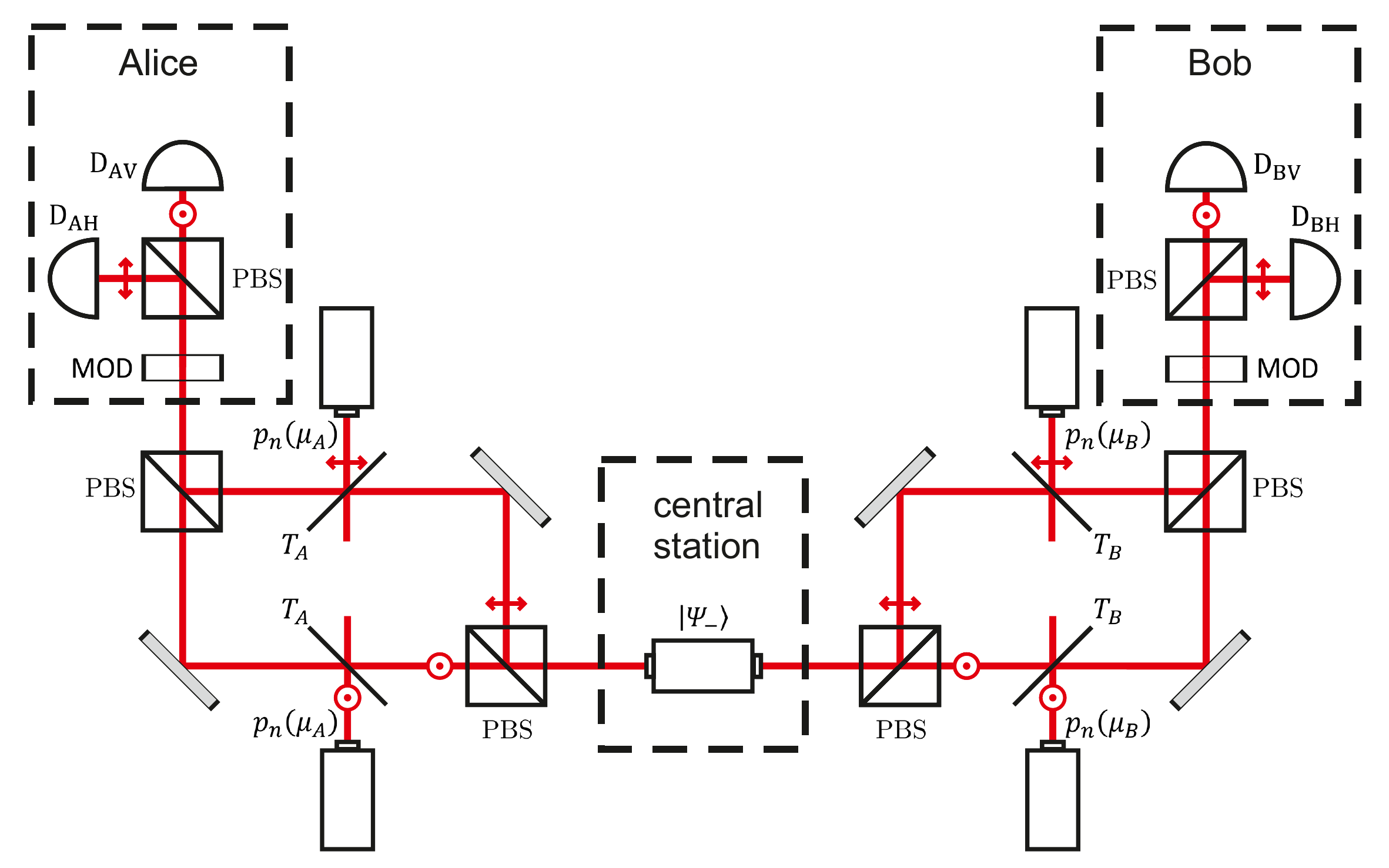} 
    \caption{DV QKD scheme with the source of entangled photon pairs located in the central station. The channels connecting it with the participants of the protocol are assumed to be lossy and noisy and their model is based on Ref.\,\cite{Lasota2017}. PBS: polarizing beam-splitter, MOD: polarization modulator.}
    \label{fig:CharlieSource}
\end{figure}

First we consider the scheme presented in Fig.\,\ref{fig:CharlieSource}, where the entangled photon pairs are generated in the central station and subsequently sent to Alice and Bob through lossy and noisy quantum channels. In order to make this figure (and also Figs.\,\ref{fig:CharlieDetectors}--\ref{fig:CharlieDetectorsCV}) more transparent we framed the laboratories of the legitimate parties and the central station with dashed black lines. Every setup element laying outside of these lines is assumed to be a part of the channel model. The measurement systems utilized by the legitimate parties are assumed to be identical, each consisting of a polarization modulator, a polarizing beam-splitter and a pair of single-photon detectors. Changing the polarization of the incoming photons in the polarization modulators is equivalent to changing the measurement bases by the trusted parties. We consider both photon-number-resolving (PNR) and binary on/off detectors used by Alice and Bob, with efficiencies $\eta_A$ and $\eta_B$, respectively. In both cases the measurement systems are assumed to be noiseless. This assumption is well-justified in our study, since we focus our attention on the scenario with relatively high transmittance of quantum channels, where the probability of observing a dark count is negligible comparing to the other detection events (with typical dark count rates ranging in kHz or below \cite{Eisaman2011} and detection windows no longer than single nanoseconds it is unlikely to expect the probability of a dark count per detection window to exceed $10^{-5}$). On the other hand we consider thermal-bath type of channel noise, coupled to the signals during their propagation from the central station to Alice's and Bob's laboratories. This noise model was previously used to compare the robustness of CV and DV QKD to the channel noise in the P\&M scenario, as it allows equivalent parametrization of the channel noise for both families of the protocols, while corresponding to single-mode thermal noise, typically observed in practical quantum communication \cite{Lasota2017}. 
	
We first consider the trusted-device scenario, in which we assume that the source produces perfect $|\Psi_-\rangle$ Bell states with probability $q$ and no photons with probability $1-q$. Each of the generated photons is then successfully collected by the further QKD setup with probability $\xi$. In order to clearly compare the DV and CV families of the protocols in their ultimate performance we focus on deterministic type of photon pair sources, which do not produce more than one photon pair in a single event. Such performance is already practically achieved by the quantum-dot type sources \cite{Huber2018,Basset2021}. Although nowadays the probabilistic sources, especially the ones based on spontaneous parametric down-conversion (SPDC) process, are still more often used in QKD implementations, recent theoretical studies suggest that their deterministic counterparts have the potential to outperform them in the future applications \cite{Hosak2021}.
	
We consider entanglement-based equivalents of the BB84 \cite{Bennett84} and six-state \cite{Bruss98} protocols utilizing photon polarization to generate the key, with asymmetric sifting procedure \cite{Lo05} used by Alice and Bob. When quantum bit error rate (QBER), $Q$, is independent of the measurement basis, which is to be expected with our assumptions on noise, the key generation rate for the BB84 protocol can be lower-bounded as follows \cite{Gottesman04,Kraus05,Renner05}:
\begin{equation}
    K^\mathrm{BB84}=p_\mathrm{exp}\max[0,1-2H(Q)],
\label{eq:Kbb84}
\end{equation}
where
\begin{equation}
    H(Q)=-Q\log_2Q-\left(1-Q\right)\log_2\left(1-Q\right)
\end{equation}
is the binary Shannon entropy. Analogously, for six-state protocol one has \cite{Renner05}:
\begin{equation}
    K^\mathrm{6state}=p_\mathrm{exp}\max[0,1-F(Q)],
\label{eq:K6state}
\end{equation}
where
\begin{equation}
    F(Q)=-\left(1-\frac{3Q}{2}\right)\log_2\left(1-\frac{3Q}{2}\right)-\frac{3Q}{2}\log_2\frac{Q}{2}.
\end{equation}
	
In order to express $p_\mathrm{exp}$ and $Q$ using experimental parameters let us first denote the probability of a given detector, belonging to the party $X$ ($A$ for Alice, $B$ for Bob), to collect $i$ noise photons in a single key generation attempt by $\pi_i(\mu_X,T_X,\eta_X)$. It is equal to
\begin{equation}
    \pi_i(\mu_X,T_X,\eta_X)=\sum_{n=i}^\infty p_n(\mu_X){n \choose i}\left[(1-T_X)\eta_X\right]^i\left[1-(1-T_X)\eta_X\right]^{n-i},
\end{equation}
where $T_X$ is the transmittance of the channel connecting the central station with $X$ and
\begin{equation}
    p_n(\mu_X)=\frac{\mu_X^n}{(1+\mu_X)^{n+1}}
\end{equation}
is the thermal statistics, which is typical for the channel noise. In the PNR detectors' case the attempt is accepted if and only if Alice and Bob have chosen identical measurement bases (the probability of which can be made arbitrarily close to one with the assumed asymmetric sifting procedure and infinitely long key) and only one photon is collected by each party. Therefore
\begin{equation}
    p_\mathrm{exp}^\mathrm{PNR}=q\,p_\mathrm{acc,1}^\mathrm{PNR}+(1-q)\,p_\mathrm{acc,0}^\mathrm{PNR},
\label{eq:pexpfull}
\end{equation}
where
\begin{equation}
    p_\mathrm{acc,1}^\mathrm{PNR}=r_A\,r_B,
\end{equation}
\begin{equation}
    r_X=\xi_XT_X\eta_X\left[\pi_0(\mu_X,T_X,\eta_X)\right]^2+2(1-\xi_XT_X\eta_X)\,\pi_0(\mu_X,T_X,\eta_X)\,\pi_1(\mu_X,T_X,\eta_X)
\end{equation}
and
\begin{equation}
    p_\mathrm{acc,0}^\mathrm{PNR}=4\,\pi_0(\mu_A,T_A,\eta_A)\,\pi_1(\mu_A,T_A,\eta_A)\,\pi_0(\mu_B,T_B,\eta_B)\,\pi_1(\mu_B,T_B,\eta_B).
\end{equation}
Among the accepted events only the ones in which both signal photons and no noise photons are detected are error-free. In all of the other cases the probability of an error is $50\%$. Thus, the QBER is given by
\begin{equation}
    Q^\mathrm{PNR}=\frac{p_\mathrm{exp}^\mathrm{PNR}-q\,\xi_AT_A\eta_A\left[\pi_0(\mu_A,T_A,\eta_A)\right]^2\xi_BT_B\eta_B\left[\pi_0(\mu_B,T_B,\eta_B)\right]^2}{2p_\mathrm{exp}^\mathrm{PNR}}.
\label{eq:Qfull}
\end{equation}
	
In order to prevent the potential eavesdropper from threatening the security of the generated key with large-pulse attacks \cite{Lutkenhaus1999} we assume here that when using on/off detectors the trusted parties discard from the key only the attempts in which at least one of them received no click at all. In case of a double click event observed by Alice or Bob, the value of the bit is chosen by the party randomly. With this assumption the formula for $p_\mathrm{exp}$ transforms into
\begin{equation}
    p_\mathrm{exp}^\mathrm{on/off}=q\,p_\mathrm{acc,1}^\mathrm{on/off}+(1-q)\,p_\mathrm{acc,0}^\mathrm{on/off},
\end{equation}
where
\begin{equation}
    p_\mathrm{acc,1}^\mathrm{on/off}=s_A\,s_B,
\end{equation}
\begin{equation}
    s_X=1-(1-\xi_XT_X\eta_X)\left[\pi_0(\mu_X,T_X,\eta_X)\right]^2
\end{equation}
and
\begin{equation}
    p_\mathrm{acc,0}^\mathrm{on/off}=\left\{1-\left[\pi_0(\mu_A,T_A,\eta_A)\right]^2\right\}\left\{1-\left[\pi_0(\mu_B,T_B,\eta_B)\right]^2\right\}.
\end{equation}
This time the only error-free events are the ones when both signal photons are detected by the trusted parties and there are no simultaneous clicks in the two remaining detectors. In all of the other accepted cases the probability of an error is once again equal to $50\%$. Thus
\begin{equation}
    Q^\mathrm{on/off}=\frac{p_\mathrm{exp}^\mathrm{on/off}-q\,\xi_AT_A\eta_A\,\pi_0(\mu_A,T_A,\eta_A)\,\xi_BT_B\eta_B\,\pi_0(\mu_B,T_B,\eta_B)}{2p_\mathrm{exp}^\mathrm{on/off}}.
\end{equation}
	
\subsection{Set of detectors in the central station}
	
\begin{figure}[tbh]
    \centering
    \includegraphics[width=\textwidth]{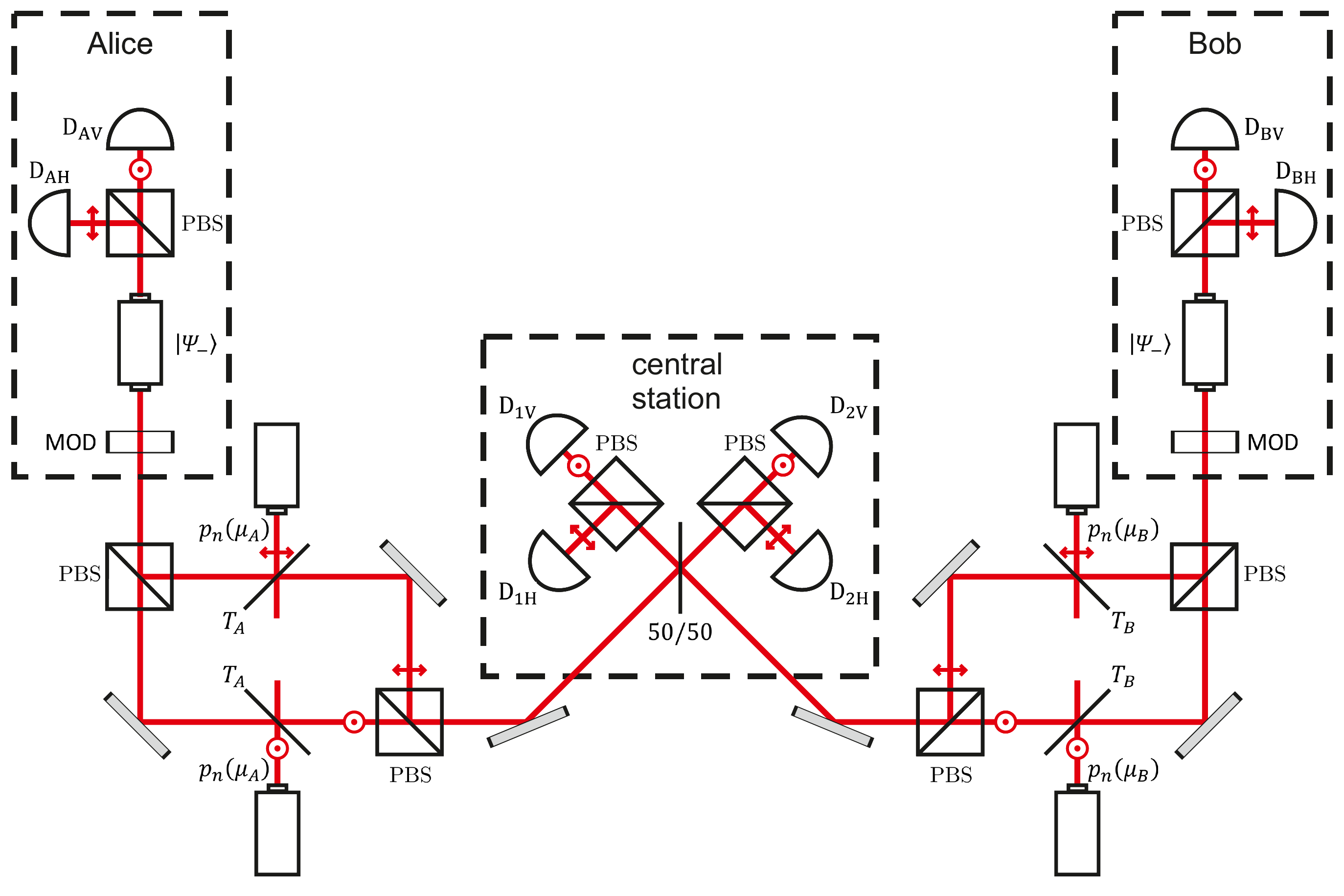} 
    \caption{DV QKD scheme with the central station performing joint detection of the pairs of photons generated by Alice and Bob, adopted from Ref.\,\cite{Kaltenbaek09}. The channel model and the abbreviations used in this picture are the same as in Fig.\,\ref{fig:CharlieSource}.}
\label{fig:CharlieDetectors}
\end{figure}
	
Alternatively, EPR-based QKD can be realized using the scheme presented in Fig.\,\ref{fig:CharlieDetectors}. Here the roles of the parties and the central station are reversed in comparison with the previous scheme. Both Alice and Bob possess a photon-pair source, producing the states $|\Psi_-\rangle$. One of the photons from each state generated by these sources is measured by the local detection system, while the other one is sent to the central station, which is supposed to jointly measure the photons received from the trusted parties and announce the result publicly. This scheme is analogous to the setup configuration utilized by Kaltenbaek \emph{et al.} in Ref.\cite{Kaltenbaek09} for the demonstration of entanglement swapping. Here the central station does not have to be trusted, since its operator cannot gain any knowledge on the specific states sent by Alice and Bob without introducing errors in their raw key. Thus, the setup configuration illustrated in Fig.\,\ref{fig:CharlieDetectors} can be used for measurement-device-independent (MDI) QKD. In fact, it is very similar to the original MDI QKD scheme \cite{Lo2012}, where the trusted parties' laboratories were just assumed to contain sources of weak coherent pulses, instead of photon-pair sources and additional detection systems as it is done here.
	
Polarization modulators placed in the paths of photons travelling to the central station allow Alice and Bob to perform MDI versions of the BB84 and six-state protocols. In the noiseless scenario if their polarization states belong to the horizontal/vertical basis coincidence clicks registered in a pair of the detectors $\mathrm{D}_{i\mathrm{H}}$ and $\mathrm{D}_{j\mathrm{V}}$, where $i,j=\left\{1,2\right\}$, unambiguously indicate that they were polarized orthogonally to each other.	Alternatively, in the diagonal/anti-diagonal and right-circular/left-circular bases the pairs of detectors $\mathrm{D}_{1\mathrm{H}}$, $\mathrm{D}_{1\mathrm{V}}$ or $\mathrm{D}_{2\mathrm{H}}$, $\mathrm{D}_{2\mathrm{V}}$ ($\mathrm{D}_{1\mathrm{H}}$, $\mathrm{D}_{2\mathrm{V}}$ or $\mathrm{D}_{2\mathrm{H}}$, $\mathrm{D}_{1\mathrm{V}}$) register coincidences only when the photons entering the central station have the same (orthogonal) polarizations. Therefore, if any of the aforementioned coincidences is announced by the central station and the trusted parties have chosen the same polarization bases, through the measurement of his own photon Bob automatically gains knowledge on the polarization of the photon kept by Alice and vice versa. While technically all the bases can be used to generate the key, we focus here on the asymmetric versions of the protocols, in which only the horizontal-vertical basis is utilized for this purpose, while the other basis (or bases) serves to detect potential eavesdropping attempts.
	
In our analysis of the performance of the setup configuration illustrated in Fig.\,\ref{fig:CharlieDetectors} we adopt the analogous assumptions for both sources, all of the detectors and the quantum channels connecting the trusted parties with the central station as in Sec.\,\ref{Sec:SourceMiddle}. In particular, the assumed absence of noise in the detectors utilized by Alice and Bob means that whenever at least one of the sources generated vacuum, the respective local measurement system would not register a click and the event would be discarded from the key. Therefore, the security analysis can be limited only to the cases when both sources generated exactly one pair of photons. The expected probability for accepting a given key generation attempt is equal to
\begin{equation}
    p_\mathrm{exp}=q_A\xi_A\eta_A\,q_B\xi_B\eta_B\sum_{i,j=1}^2\left[P_\mathrm{HV}^{i\mathrm{H},j\mathrm{V}}+P_\mathrm{VH}^{i\mathrm{H},j\mathrm{V}}+P_\mathrm{HH}^{i\mathrm{H},j\mathrm{V}}+P_\mathrm{VV}^{i\mathrm{H},j\mathrm{V}}\right],
\label{eq:pexpMDIproper}
\end{equation}
where $q_X$ and $\xi_X$ denote, respectively, the probability for generating a pair of photons and the photon collection efficiency for the source owned by the party $X$, while $P_\mathrm{YZ}^{i\mathrm{H},j\mathrm{V}}$ is the joint probability for Alice to detect a $\mathrm{Y}$-polarized photon, Bob to detect a $\mathrm{Z}$-polarized photon and the central station to detect a coincidence in the detectors $\mathrm{D}_{i\mathrm{H}}$ and $\mathrm{D}_{j\mathrm{V}}$. While the first two probabilities inside the square bracket in the formula (\ref{eq:pexpMDIproper}) describe the events providing Alice and Bob with matching key bits, the other two correspond to errors. Therefore, QBER can be calculated as
\begin{equation}
    Q=q_A\xi_A\eta_A\,q_B\xi_B\eta_B\sum_{i,j=1}^2\left[P_\mathrm{HH}^{i\mathrm{H},j\mathrm{V}}+P_\mathrm{VV}^{i\mathrm{H},j\mathrm{V}}\right]/p_\mathrm{exp}.
\label{eq:qberMDIproper}
\end{equation}
	
Although in general the error rate registered in the diagonal/anti-diagonal or right-circular/left-circular bases is given by a different formula,
\begin{equation}
Q_\mathrm{other\, bases}=q_A\xi_A\eta_A\,q_B\xi_B\eta_B\left[\sum_{i=1}^2\left(P_\mathrm{HV}^{i\mathrm{H},i\mathrm{V}}+P_\mathrm{VH}^{i\mathrm{H},i\mathrm{V}}\right)+\sum_{i\neq j}\left(P_\mathrm{HH}^{i\mathrm{H},j\mathrm{V}}+P_\mathrm{VV}^{i\mathrm{H},j\mathrm{V}}\right)\right]/p_\mathrm{exp}.
\end{equation}
in the case of the assumed symmetry of the setup and the uniformity of the channel noise, we obtain $Q_\mathrm{other\, bases}=Q$. Furthermore, the formulas (\ref{eq:pexpMDIproper}) and (\ref{eq:qberMDIproper}) can be simplified to
\begin{equation}
    p_\mathrm{exp}=8\,q_A\xi_A\eta_A\,q_B\xi_B\eta_B\left[P_\mathrm{HV}^{1\mathrm{H},1\mathrm{V}}+P_\mathrm{HH}^{1\mathrm{H},1\mathrm{V}}\right].
\label{eq:pexpMDI}
\end{equation}
and
\begin{equation}
    Q=8\,q_A\xi_A\eta_A\,q_B\xi_B\eta_BP_\mathrm{HH}^{1\mathrm{H},1\mathrm{V}}/p_\mathrm{exp},
\label{eq:qberMDI}
\end{equation}
respectively. In the case when the trusted parties perform the analog of the BB84 [six-state] protocol using the setup configuration presented in Fig.\,\ref{fig:CharlieDetectors}, the results of Eq.\,(\ref{eq:pexpMDIproper}) and (\ref{eq:qberMDIproper}) should be inserted into Eq.\,(\ref{eq:Kbb84}) [Eq.\,(\ref{eq:K6state})] in order to calculate the lower bound for the key generation rate. The exact expressions for $P_\mathrm{HV}^{1\mathrm{H},1\mathrm{V}}$  and $P_\mathrm{HH}^{1\mathrm{H},1\mathrm{V}}$ in the analyzed scenario, both for the cases when the PNR and on/off detectors are utilized, are provided in the Appendix 1.
	
\section{Security of EPR-based CV QKD}
\label{Sec:CVtheory}
    
Contrary to DV QKD, described in the previous section, which is based on direct photodetection, CV QKD uses homodyne detection of quadrature observables, performed by the trusted parties. Nevertheless, the principles of security analysis remain the same and rely on assessment of the lower bound on the secure key rate as an information advantage of the trusted parties over the upper bound on the information accessible to an eavesdropper:
\begin{equation}
    K^{CV}=\max\left[0, I_{AB}-\chi_{EA/EB}\right].
\label{eq:keyCV}
\end{equation}
Here $I_{AB}$ is the classical mutual information between the parties and $\chi_{EA/EB}$ denotes the Holevo bound, constraining the information between the eavesdropper and the reference side of the protocol, either Alice (A) or Bob (B). We consider Gaussian CV QKD and analyze security against optimal Gaussian collective attacks \cite{Navascues2006,GarciaPatron2006}, which can be extended to finite-size regime \cite{Leverrier2010} and general (coherent) attacks \cite{Leverrier2013} using de Finetti reduction \cite{Leverrier2017}. Note that the CV prepare-and-measure protocols can also be implemented using discrete, generally non-Gaussian modulation \cite{Ghorai2019}, which can simplify the error correction procedure, but requires complex optimizations for security analysis. In the entanglement-based schemes, considered in this work, the states are all Gaussian and the respective Gaussian security analysis therefore applies.
   	
For this analysis we assume the most typical CV Gaussian entangled states, namely the two-mode squeezed vacuum (TMSV) states \cite{Weedbrook2012}. Those states are maximally entangled, hence providing the ultimate performance of EPR-based CV QKD, and are fully described by mean values and second moments of quadrature operators, defined for a given mode as $\hat{x}=\hat{a}^\dagger+\hat{a}$ and $\hat{p}=i(\hat{a}^\dagger-\hat{a})$. Further, with no loss of generality, we assume that x-quadrature is measured by the trusted parties to extract the key bits. Then, for evaluation of the mutual information $I_{AB}$ it is sufficient to know the x-quadrature covariance matrix $\gamma^{(x)}_{AB}=\left(\begin{array}{cc}
    V_A & C_{AB} \\
    C_{BA} & V_B \\
\end{array}\right)$ of the state shared between Alice and Bob, which contains variances $V_A=\langle x_A^2 \rangle$, $V_B=\langle x_B^2 \rangle$ and correlation $C_{AB}=\langle x_Ax_B \rangle$ between the quadratures $x_A$ and $x_B$ measured by Alice and Bob respectively (taking into account their zero mean values for TMSV states). We then evaluate the mutual information $I_{AB}=(1/2)\log_2{(V_A/V_{A|B})}$, where conditional variance $V_{A|B}=V_A-C_{AB}^2/V_B$. 
	
The Holevo bound, on the other hand, is obtained in the most general assumption that Eve holds the purification of the state shared by Alice and Bob after its propagation through the untrusted noisy channels. In the special case of the central station placed exactly in the middle of the distance separating Alice and Bob, which we analyze further, the Holevo bound is the same for either party being the reference side of the protocol (\emph{e.g.}, Bob) and is then obtained as $\chi_{EB}=S(E)-S(E|B)$. Here  $S(E)$ denotes the von Neumann entropy of the state collectively measured by Eve and $S(E|B)$ is the von Neumann entropy of Eve's state conditioned on Bob's measurement. As the TMSV is initially pure and Eve holds the purification of the noise added in the channel, the triangle inequality \cite{Araki1970} implies that $S(E)=S(AB)$ and, after Bob's projective measurement, $S(E|B)=S(A|B)$. For the Gaussian states, $S(AB)=\sum_i \text{G } (\frac{\lambda_i-1}{2})$, where $\lambda_i$ are symplectic eigenvalues of the overall covariance matrix $\gamma_{AB}$ of the elements $\gamma_{ij}=\langle r_ir_j\rangle$, $r_{i,j}\in\{x_A,p_A,x_B,p_B\}$ being one of the quadrature observables in the modes $A$ or $B$, measured respectively by Alice and Bob, and $\text{G}(x)=(x+1)\log_2(x+1)-x\log_2 x$ in the bosonic entropy function \cite{Weedbrook2013}. Similarly, $S(A|B)=\text{G } (\frac{\lambda_3-1}{2})$, where $\lambda_3$ is the symplectic eigenvalue of the conditional covariance matrix $\gamma_{A|B}$ after Bob's measurement of x-quadrature. The conditional covariance matrix is obtained as $\gamma_{A|B}=\gamma_A-\sigma_{AB}(X\gamma_BX)^{MP}\sigma_{AB}^T$, where $\gamma_A$, $\gamma_B$, and $\sigma_{AB}$ are the modes' and correlation submatrices of the overall matrix $\gamma_{AB}=
\left(
\begin{array}{cc}
    \gamma_A & \sigma_{AB} \\
    \sigma_{AB} &  \gamma_B \\
\end{array}
\right)$, $X=
\left(
\begin{array}{cc}
    1 & 0 \\
    0 &  0 \\
\end{array}
\right)$ stands for x-quadrature measurement, $MP$ is the Moore-Penrose inverse of a matrix. This defines the framework for security analysis of a perfectly implemented CV QKD protocol.
       
\subsection{Source of entanglement in the central station}
 
\begin{figure}[tbh]
    \centering
    \includegraphics[width=0.8\textwidth]{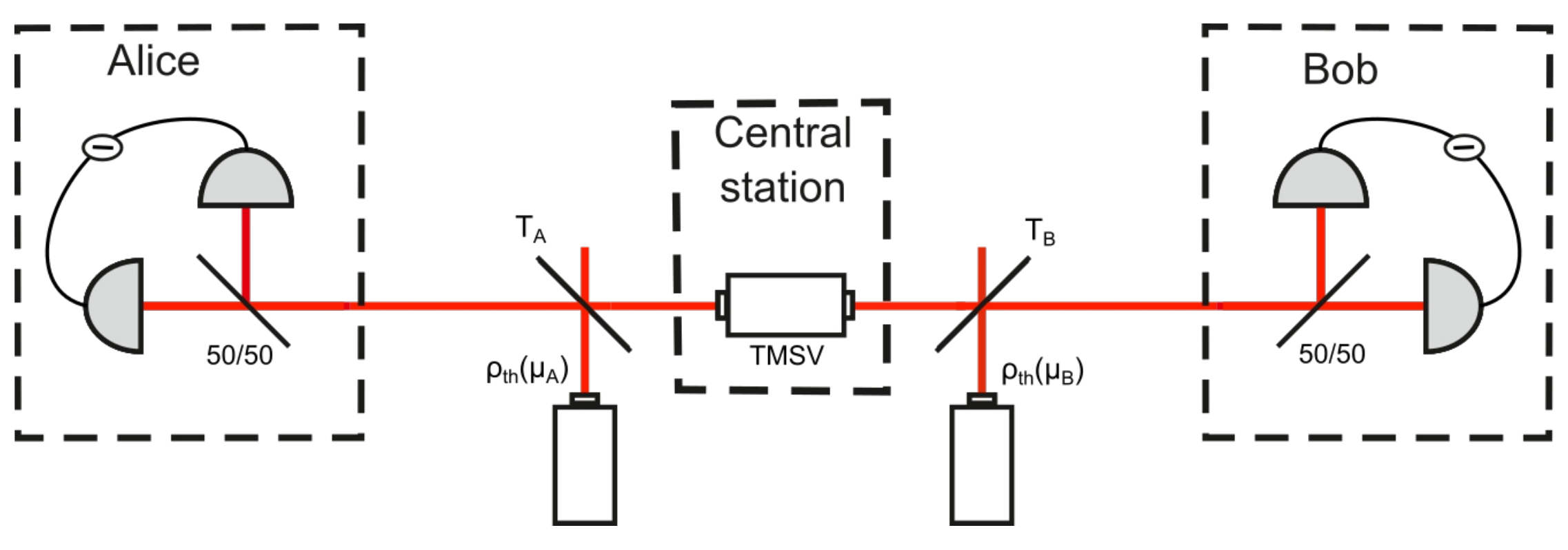} 
    \caption{The CV analogue of the QKD scheme with the source of entanglement placed in the central station, introduced in Fig.\,\ref{fig:CharlieSource}. }
\label{fig:CharlieSourceCV}
\end{figure}

The CV version of the QKD setup configuration with the central station containing source of the entangled states is illustrated in Fig.\,\ref{fig:CharlieSourceCV}. The state produced by the source (which does not have to be trusted and can be identified with Eve) is subsequently shared between Alice and Bob who both perform homodyne measurement of one of the quadratures (we assume that x-quadrature is measured) using homodyne detectors. After the end of transmission they perform error correction and privacy amplification on the accumulated data to obtain the secret key.
	
The initial TMSV state with quadrature variance $V$ is described by the covariance matrix of the form
\begin{equation}\gamma_{AB}=
    \left(
    \begin{matrix}
	V ~\mathbb{I}& \sqrt{V^2-1} ~\mathbb{Z}\\
	\sqrt{V^2-1} ~\mathbb{Z} &  V ~\mathbb{I} \\
    \end{matrix}
    \right),
\label{eq:covmat}
\end{equation}
where $\mathbb{I}=\left(
\begin{array}{cc}
    1 & 0 \\
    0 & 1 \\
\end{array}
\right)$ and $\mathbb{Z}=\left(
\begin{array}{cc}
    1 & 0 \\
    0 & -1 \\
\end{array}
\right)$. The noise added to the state while passing through the channels of transmittance $T$ is modelled as a thermal bath, similarly to the DV case, studied in the previous Sections. It can be then equivalently described as each mode of TMSV being coupled to a thermal state on a beam-splitter $T$ (as it was illustrated in Fig.\,\ref{fig:CharlieSourceCV}). The thermal state with mean photon number $\mu_{A/B}$ is described by the covariance matrix $\gamma_{th_{A/B}} = \left(
\begin{matrix}
    2 \mu_{A/B} +1 & 0\\
    0 &  2\mu_{A/B} +1 \\
\end{matrix}
\right)=\left(
\begin{matrix}
    N_{A/B} & 0\\
    0 & N_{A/B} \\
\end{matrix}
\right)$, where $N=2\mu+1$ is its quadrature variance. Then, the covariance matrix of the shared state is modified by the channels to
\begin{equation}
\small{\gamma_{AB}= \left(
    \begin{array}{cccc}
        T_A V+(1-T_A) N_{A} & 0 & \sqrt{T_A} \sqrt{T_B} \sqrt{V^2-1} & 0 \\
	0 & T_A V+(1-T_A) N_{A} & 0 & -\sqrt{T_A}              \sqrt{T_B} \sqrt{V^2-1} \\
	\sqrt{T_A} \sqrt{T_B} \sqrt{V^2-1} & 0 & T_B V+(1-     T_B) N_{B}  & 0 \\
	0 & -\sqrt{T_A} \sqrt{T_B} \sqrt{V^2-1} & 0 & T_B      V+(1-T_B) N_{B}  \\
    \end{array}
\right)}.
\label{eq:covmat_source}
\end{equation}
The covariance matrix of the conditional state after Bob's measurement then reads	
\begin{equation}
\gamma_{A|B}= \left(
    \begin{array}{cc}
        T_A V+(1-T_A) N_{A} -\frac{T_A T_B \left(V^2-1\right)}{T_B V+(1-T_B) N_B}& 0 \\
	0 & T_A V+(1-T_A) N_{A} \\
    \end{array}
\right).
\label{eq:covmat_sourcecond}
\end{equation}
	
Assuming the same transmittance values $T$ in both channels, the key rate (\ref{eq:keyCV}) takes the analytical expression
\begin{equation}
    \begin{split}
	K^{CV}&=  \frac{1}{2}\log_2 \left(\frac{(N_A (1-       T)+T V) (N_B (1-T)+T V)}{(N_A (1-T)+T V) (N_B (1-      T)+T V)-T^2 \left(V^2-1\right)}\right)-                G(\frac{\lambda_1-1}{2})\\&-G(\frac{\lambda_2-1}       {2})+G(\frac{\lambda_3-1}{2}),
    \end{split}
\end{equation}
where
\begin{equation}
    \begin{split}
	\lambda_{1,2}&=\left[\frac{(1-T)^2                     \left(N_A^2+N_B^2\right)+2 T (1-T) V (N_A+N_B)+2       T^2}{2}\right.\\&\left.\pm \frac{(1-T) (N_A-N_B)       \sqrt{(1-T)^2 (N_A+N_B)^2+4 T^2+4 T (1-T) V            (N_A+N_B)}}{2}\right]^{1/2}
    \end{split}
\end{equation}
are the symplectic eigenvalues of $\gamma_{AB}$, given by Eq.\,(\ref{eq:covmat_source}) and 
\begin{equation}
    \lambda_3=\frac{\sqrt{N_A (1-T)+T V} \sqrt{N_A (1-T) (N_B (1-T)+T V)+T (N_B (1-T) V+T)}}{\sqrt{N_B (1-T)+T V}}
\end{equation} 
is the  symplectic eigenvalue of $\gamma_{A|B}$, given by Eq.\,(\ref{eq:covmat_sourcecond}).
Adding assumption that mean photon number $\mu$ is the same for both thermal baths the expression for the key simplifies to
\begin{equation}
    \begin{split}
 	K^{CV}&=  \frac{1}{2} \log_2\frac{(T V + (1 - T)       N)^2}{(T V + (1 - T) N)^2-T^2 ( 
        V^2-1)}\\&-\frac{1}{2} \log_2 \frac{1}{4} \left[ (T V + (1 - T) N)^2-T^2 ( 
        V^2-1) -1\right] \\
  	&-\frac{1}{2}  \sqrt{(T V + (1 - T) N)^2-T^2 ( 
        V^2-1)} \log_2 \frac{\sqrt{(T V + (1 - T) N)^2-T^2 (V^2-1)}+1}{\sqrt{(T V + (1 - T) N)^2-T^2 ( 
        V^2-1)}-1}. 
    \end{split}
\label{eq:KCVsource}
\end{equation}
In the limit of infinitely strong entanglement of the source, $V\rightarrow\infty$, the expression for the key rate tends to $K^{CV} \rightarrow  \log_2 \frac{1}{N(1-T)}-\frac{1}{\ln 2}$, giving a threshold for maximal tolerable variance of the thermal noise $N=\frac{1}{e (1-T)}$, or, equivalently, tolerable quadrature excess noise with respect to the channel input (typical parametrization used in CV QKD) $\varepsilon_{max}=1+\frac{1-e}{e T}$. The transmittance of each of the symmetric channels for ideal CV QKD with entanglement in the middle is bounded by $T_{min}=1-\frac{1}{e N}$. For pure loss channel ($\mu=0$) one can obtain $T_{min}\approx 0.63$, which corresponds to approx.\,10 km of standard telecom fiber with attenuation of -0.2 dB/km. The protocol cannot be implemented at lower values of channel transmittance.

\subsection{Set of detectors in the central station}

\begin{figure}[tbh]
    \centering
    \includegraphics[width=\textwidth]{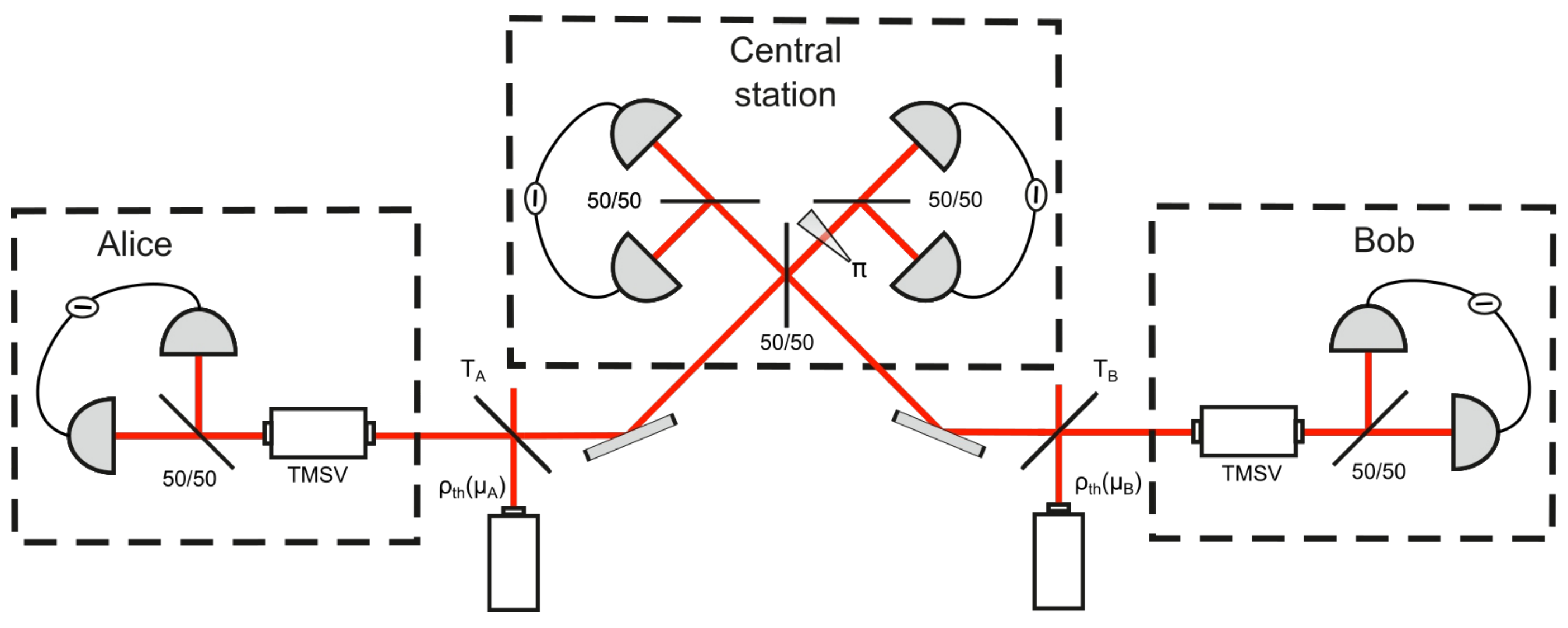} 
    \caption{The CV analogue of the QKD scheme with the set of detectors placed in the central station, introduced in Fig.\,\ref{fig:CharlieDetectors}. Here the central station performs joint measurement of the opposite quadratures of the states generated by Alice and Bob ($\pi$ on one of the beam-splitter's outputs denotes a phase shift). }
\label{fig:CharlieDetectorsCV}
\end{figure}

Next we consider MDI CV QKD protocol \cite{Pirandola2015}, performed with the setup configuration shown in Fig.\,\ref{fig:CharlieDetectorsCV}. In this protocol both Alice and Bob posses their own sources to generate TMSV states. Each of them keeps one of the TMSV modes for the local homodyne detection, while sending the other mode to the central station, which performs Bell-type measurement. For this purpose the received signals are first mixed on a balanced beam-splitter and then the opposite quadratures ($\hat{x}$ and $\hat{p}$) are measured on its outputs. The central station publicly announces the measurement results $x_C$ and $p_C$, allowing the trusted parties to obtain conditional data sequences and, after error correction and privacy amplification, generate the secure key from those.

Similarly to the conventional entanglement-in-the-middle scheme, each of the trusted parties initially generates a TMSV state (\ref{eq:covmat}). One of the modes of each of those states is send to the central station through the channels with transmittance $T_A$ and $T_B$ respectively, where the modes get coupled to a thermal state with the variance $N_{A/B}=2\mu_{A/B}+1$. Announcement of the outcome of the Bell-type measurement in the central station leads to the creation of the conditional state of the remaining two modes shared between Alice and Bob, described by the following covariance matrix: 
\begin{equation}
\small{
    \gamma_{AB}^{(MDI)}=   \left(
    \begin{array}{cc}
        V-\frac{T_A \left(V^2-1\right)}{V (T_A+T_B)+(1-    T_A) N_{A}+(1-T_B) N_{B}} ~\mathbb{I}& -\frac{\sqrt{T_A} \sqrt{T_B} \left(V^2-1\right)}{T_A (V-N_{A})+T_B (V-N_{B})+N_{A}+N_{B}} ~\mathbb{Z} \\
 	-\frac{\sqrt{T_A} \sqrt{T_B} \left(V^2-1\right)}       {T_A (V-N_{A})+T_B (V-N_{B})+N_{A}+N_{B}}      
        ~\mathbb{Z} & V-\frac{T_B \left(V^2-1\right)}{V (T_A+T_B)+(1-T_A) N_{A}+(1-T_B) N_{B}} ~\mathbb{I} \\
    \end{array}
    \right)
    }.
\label{eq:covmat_det}
\end{equation}
The conditional covariance matrix after the homodyne measurement of x-quadrature at Bob's side then takes the form
\begin{equation}
    \gamma_{A|B}^{(MDI)}=  \left(
    \begin{array}{cc}
 	\frac{V [T_A+T_B+V (N_A+N_B-N_B T_B-N_A T_A )]}{V      (N_A (1-T_A)+N_B(1-T_B)+T_A V)+T_B} & 0 \\
 	0 & V-\frac{T_A \left(V^2-1\right)}{V (T_A+T_B)+      (1-T_A) N_{A}+(1-T_B) N_{B}} \\
    \end{array}
    \right).
\end{equation}

Assuming the realization of the protocol is fully symmetric, both in terms of the channel transmittance $T$ and the mean photon number $\mu$ of the thermal baths coupled to each state, either of the sides can be equivalently used as the reference for data reconciliation. In this case the secure key rate (\ref{eq:keyCV}) can be analytically expressed as	
\begin{equation}
    \begin{split}
        K^{CV}=\frac{1}{2}\log_2 \frac{\left[T(1+V^2)+2(1-T)V N \right]^2}{V [N (1-T) V+T] [N (1-T)+T V]}-\frac{1}{2}\log_2 \frac{N (1-T) \left(V^2-1\right)}{N(1-T)+T V} \\+ \frac{\sqrt{V} \sqrt{N (1-T) V+T } }{2 \sqrt{N (1-T)+T V}}\log_2 \frac{\sqrt{V} \sqrt{N (1-T) V+T }-\sqrt{N (1-T)+T V}}{\sqrt{V} \sqrt{N (1-T) V+T }+\sqrt{N (1-T)+T V}}
    \label{eq:KCVMDI}
    \end{split},
\end{equation}
with $N=2 \mu+1$. Similarly to the standard entanglement-based protocol, discussed in the previous subsection, in the limit of infinitely strong entanglement of the source, $V\rightarrow\infty$, the secret key rate for MDI CV QKD tends to $K^{CV} \rightarrow \log_2 \frac{T}{N(1-T)}-\frac{1}{\ln 2}$. Hence, the threshold for maximal tolerable variance of the thermal noise is $N=\frac{T}{e (1-T)}$, or, in terms of the quadrature excess noise with respect to the channel input, $\varepsilon_{max}=1+\frac{T-e}{e T}$.  The secure transmittance of symmetric channels for ideal MDI CV QKD is bounded by $T_{min}=\frac{e N}{e N +1}$, with  $T_{min}\approx0.73$ for pure loss channel ($\mu=0$). This corresponds to less than 7 km of the standard telecom fiber, meaning that the protocol is even more sensitive to the channel attenuation, than the standard EPR-based protocol described before. The reason for this behavior is in the fact that after the Bell-type measurement and announcement of the outcomes in the MDI protocol, the resulting entangled state, shared between Alice and Bob, is less entangled than the initial entangled states at Alice's and Bob's stations. This can be quantified, \emph{e.g.}, using Gaussian logarithmic negativity \cite{Serafini2004}, which, in the limit of infinite initial entanglement in pure loss channel, reads $-\log_2{(1-T)}$ for the standard EPR-based scheme and $\log_2{T}-\log_2{(1-T)}$ for the MDI scheme, see Appendix 2 for details. The latter is evidently more sensitive to loss, which also affects the robustness of the respective QKD protocol.
	
\section{Robustness of the EPR-based schemes against channel noise}
\label{Sec:Results}
	
\begin{figure}[tbh]
    \centering
    \includegraphics[width=1.0\textwidth]{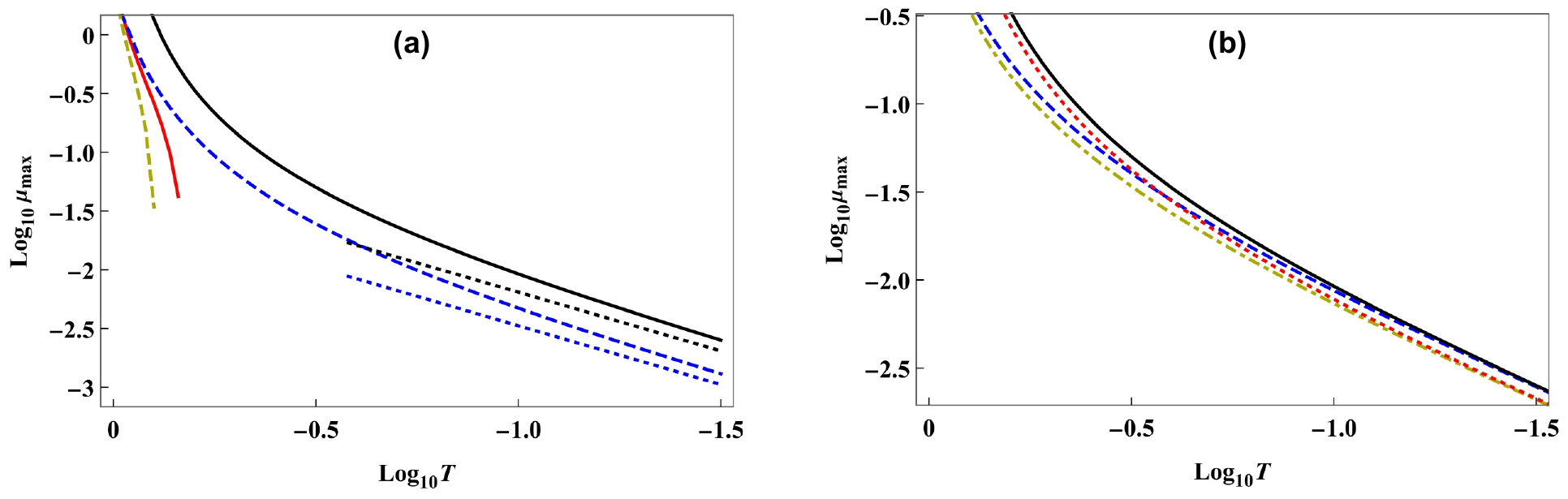} 
    \caption{(a) Maximal values of the channel noise $\mu$ for which it is possible to generate the secure key, plotted as a function of channel transmittance $T$, calculated numerically for the cases of Alice and Bob utilizing entanglement-based version of the six-state protocol in the setup configuration shown in Fig.\,\ref{fig:CharlieSource} (black solid line), MDI six-state protocol in the setup configuration shown in Fig.\,\ref{fig:CharlieDetectors} (blue dashed line), CV QKD protocol in the setup configuration shown in Fig.\,\ref{fig:CharlieSourceCV} (red solid line) and MDI CV QKD protocol in the setup configuration shown in Fig.\,\ref{fig:CharlieDetectorsCV} (yellow dashed line). All the plots were made with the assumption that the detection systems used to implement these protocols are ideal. Additionally, the black [blue] dotted line denotes the analytical approximation of the function $\mu_\mathrm{max}(T)$ for the six-state protocol realized with the Fig.\,\ref{fig:CharlieSource} [Fig.\,\ref{fig:CharlieDetectors}] scheme, given by the Eq.\,(\ref{eq:mu}) [Eq.\,(\ref{eq:mu2})], valid for $T\ll1$. (b) Analogous comparison of the performance of the entanglement-based version of the six-state [BB84] protocol, realized in the setup configuration shown in Fig.\,\ref{fig:CharlieSource}, with PNR detectors (black solid [red dotted] line) or binary on/off detectors (blue dashed [yellow dot-dashed] line) with detection efficiency $\eta=100\%$. The results in both panels were obtained with the assumption that all the sources utilized for DV and CV protocols are ideal.}
\label{fig:MainPlotsIdealSources}
\end{figure}

\begin{figure}[tbh]
    \centering
    \includegraphics[width=1.0\textwidth]{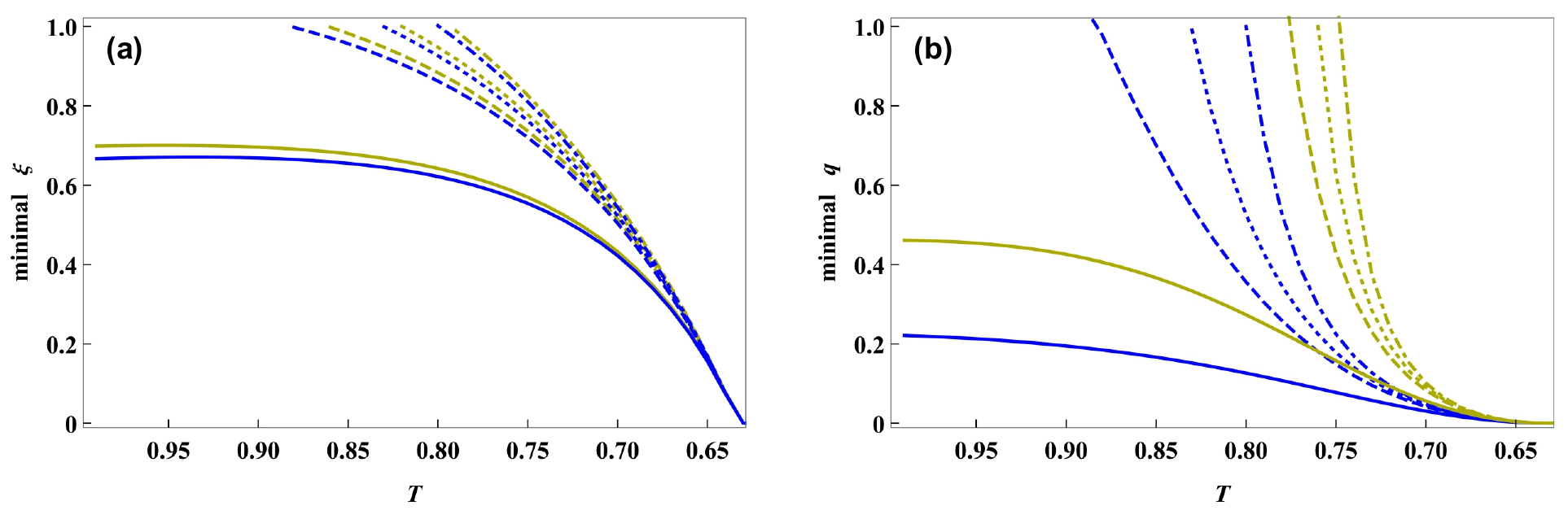} 
    \caption{Minimal value of (a) single photon collection efficiency $\xi$, (b) photon-pair generation probability $q$ needed for the entanglement-based version of the six-state protocol, realized with the setup configuration pictured in Fig.\,\ref{fig:CharlieSource}, to remain more resilient to the channel noise than the corresponding CV QKD protocol performed with ideal setup elements, plotted as a function of the channel transmittance $T$ for the case of ideal PNR detectors (solid lines) and binary on/off detectors with efficiency $\eta=100\%$ (dashed lines), $\eta=80\%$ (dotted lines) and $\eta=60\%$ (dot-dashed lines). In panel (a) the blue (yellow) lines correspond to $q=1$ ($q=0.8$). In panel (b) the blue (yellow) lines correspond to $\xi=100\%$ ($\xi=80\%$).}
\label{fig:FigureWithDoubleClicks}
\end{figure}

In this section we compare the performance of different protocols described above in the symmetric versions of the setup configurations illustrated in Figs.\,\ref{fig:CharlieSource}--\ref{fig:CharlieDetectorsCV}. This means, in particular, that $T_A=T_B\equiv T$ and $\mu_A=\mu_B\equiv\mu$. Additionally all the single-photon detectors in the DV setup configurations are assumed to have the same detection efficiency $\eta$ and the photon-pair sources belonging to the trusted parties in Fig.\,\ref{fig:CharlieDetectors} are described by $q_A=q_B\equiv q$ and $\xi_A=\xi_B\equiv \xi$.
	
Comparison between the robustness of the EPR-based six-state and CV QKD protocols in different setup configurations can be seen in Fig.\,\ref{fig:MainPlotsIdealSources} (a). While in the case of EPR-based six-state protocol the dependence of the maximal secure value of $\mu$ on channels' transmittance turns out to be qualitatively similar to the case of regular prepare-and-measure QKD \cite{Lasota2017}, the CV QKD protocol becomes insecure when the value of $T$ is still relatively high, even when the channels are noiseless. It happens because of the fundamental 3 dB limit on the loss on the reference side in the CV QKD. One can also conclude that the source-in-the-middle type of the setup configuration presented in Fig.\,\ref{fig:CharlieSource} (Fig.\,\ref{fig:CharlieSourceCV}) is slightly more robust to the channel noise than the detectors-in-the-middle type from Fig.\,\ref{fig:CharlieDetectors} (Fig.\,\ref{fig:CharlieDetectorsCV}). However, it should be also pointed out that the security level offered by the detectors-in-the-middle setup configuration is in fact stronger than in the source-in-the-middle case since the measurement setup placed in the central station is allowed to be untrusted. To make the investigation of entanglement-based QKD schemes more complete, in the Appendix 3 we also present the comparison between the two CV setups analyzed in Sec.\,\ref{Sec:CVtheory} and the device-independent (DI) DV QKD protocol introduced in Ref.\cite{Acin07}, performed with the setup configuration shown in Fig.\,\ref{fig:CharlieSource}. 
	
Although in general it is impossible to find analytical formulas for the $\mu_\mathrm{max}(T)$ curves shown in Fig.\,\ref{fig:MainPlotsIdealSources} (a), their linear approximations can be found in the DV case when $T\ll1$. In this regime the security of the investigated schemes also requires that $\mu\ll1$. With these assumptions, if the utilized sources of photon pairs and the detectors are all ideal, the formula (\ref{eq:pexpfull}) simplifies to
\begin{equation}
    Q^\mathrm{T\rightarrow0}\approx\frac{2\mu(T+\mu)}{(T+2\mu)^2}.
\end{equation}
The maximal secure value of $\mu$ for a particular DV QKD protocol performed with the scheme presented in Fig.\ref{fig:CharlieSource} can then be estimated by taking $Q^\mathrm{T\rightarrow0}=Q_\mathrm{th}$, where $Q_\mathrm{th}$ is the appropriate QBER threshold value, and solving the above equation. The result is:
\begin{equation}
    \mu_\mathrm{max}^{T\rightarrow0}\approx\frac{T\left(2Q_\mathrm{th}+\sqrt{1-2Q_\mathrm{th}}-1\right)}{2(1-Q_\mathrm{th})}.
\label{eq:mu}
\end{equation}
This function, calculated for the six-state protocol, in which case $Q_\mathrm{th}\approx0.127$ \cite{Kraus05}, is illustrated with a black dotted line in Fig.\,\ref{fig:MainPlotsIdealSources} (a). In the same picture a blue dotted line denotes analogous analytical approximation of the $\mu_\mathrm{max}(T)$ curve for the same protocol realized with the scheme presented in Fig.\,\ref{fig:CharlieDetectors}. In this case the QBER formula (\ref{eq:qberMDI}) can be simplified to 
\begin{equation}
    Q^\mathrm{T\rightarrow0}\approx\frac{4\mu(T+\mu)}{T^2+8T\mu+8\mu^2}
\end{equation}
and $\mu_\mathrm{max}^{T\rightarrow0}$ reads
\begin{equation}
    \mu_\mathrm{max}^{T\rightarrow0}\approx\frac{T\left(2Q_\mathrm{th}+\sqrt{1-3Q_\mathrm{th}+2Q_\mathrm{th}^2}-1\right)}{2(1-Q_\mathrm{th})}.
\label{eq:mu2}
\end{equation}

The reason for choosing the six-state protocols to represent the DV QKD family in the DV vs.\,CV comparison presented in Fig.\,\ref{fig:MainPlotsIdealSources} (a) is their general superiority over the more popular four-state (BB84-analogous) protocols in terms of the resistance to the channel noise. In the case of the source-in-the-middle setup configuration it can be seen in Fig.\,\ref{fig:MainPlotsIdealSources} (b), where the maximal tolerable rate of the channel noise is plotted as a function of the channel transmittance for the entanglement-based versions of both six-state and BB84 protocols. Additionally, the panel represents comparison between the situations when the trusted parties use either the ideal PNR detectors or the binary on-off detectors with $100\%$ efficiency to measure the photons generated in the central station. Basing on the presented results it can be concluded that the ability to resolve the number of incoming photons can significantly improve the robustness of both DV protocols analyzed here to the channel noise in the regime of high channel transmittance. This effect obviously diminishes with $T\rightarrow0$, when the acceptable level of the channel noise decreases and the probability for having more than two photons entering any of the detectors becomes smaller. A similar comparison between different protocols and detection systems can be also made for the detectors-in-the-middle setup configuration, with the results completely analogous to the ones presented in Fig.\,\ref{fig:MainPlotsIdealSources} (b), as can be seen in Fig.\,\ref{fig:AdditionalFigure} (a) in Appendix 4.
	
While in the case of ideal sources and detectors the advantage of DV QKD protocols over their CV counterparts is clear for the two types of setup configurations studied in this paper, it is important to analyze how good an imperfect DV setup has to be in order to remain more resilient to the channel noise than even the perfect CV scheme. In Fig.\,\ref{fig:FigureWithDoubleClicks} we answer this question for the source-in-the-middle setup configuration. It turns out that when $T\approx1$ the six-state protocol can outperform the CV QKD protocol realized with ideal setup elements only when the trusted parties use PNR detectors and sufficiently high-standard photon-pair source. Meanwhile, for on/off detectors the maximal channel transmittance allowing this possibility is about $88\%$, even when the source is ideal. In the opposite case, when the collection efficiency $\xi$ and the photon-pair generation rate $q$ are lower than unity, this value is further reduced. Comparing the panels (a) and (b) one can conclude that from the perspective of the analyzed QKD application high collection efficiency seems to be more important for a realistic photon-pair source than high generation probability. When $T>0.8$, $\xi>60\%$ is required to beat the ideal CV QKD protocol, even when all the other setup elements are perfect, as can be seen in Fig.\,\ref{fig:FigureWithDoubleClicks} (a). Such requirement is very demanding for the currently existing deterministic photon-pair sources \cite{Schimpf2021}. However, it has already been fulfilled by some specifically-designed sources based on quantum dots \cite{Chen2018}--\cite{Wang2019}. Moreover, change in the collection efficiency heavily influences the required photon-pair generation probability, as illustrated in Fig.\,\ref{fig:FigureWithDoubleClicks} (b). On the other hand, photon-pair generation probability has much weaker influence on the required level of photon collection efficiency and $q\approx0.23$ is already sufficient for the six-state protocol to beat the performance of the ideal CV QKD scheme if the setup illustrated in Fig.\,\ref{fig:CharlieSource} is otherwise perfect. Much higher level of photon-pair generation probability, even exceeding $0.9$, has already been achieved in practice \cite{Basset2021}.
	
\begin{figure}[tbh]
    \centering
    \includegraphics[width=\textwidth]{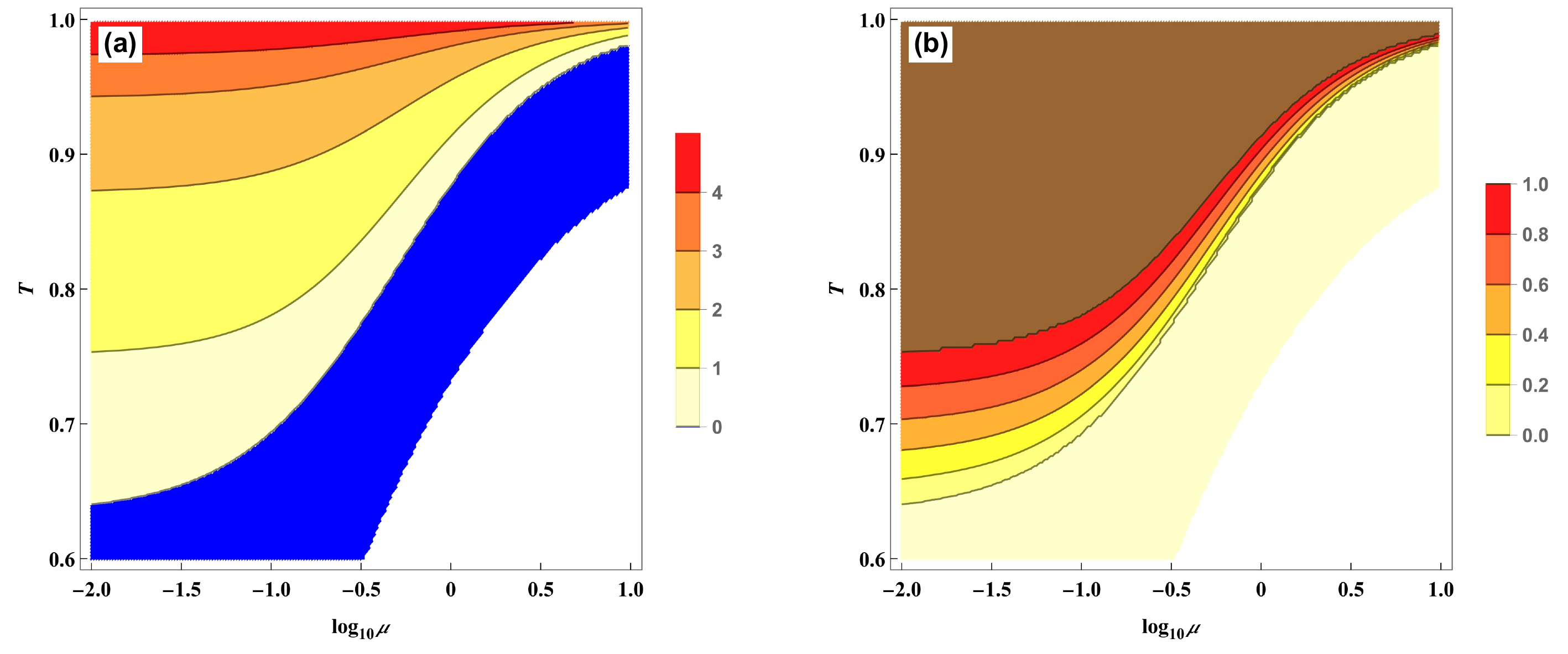} 
    \caption{(a) The ratio between the key generation rates that can be obtained using the CV QKD protocol in the setup configuration presented in Fig.\,\ref{fig:CharlieSourceCV} and the entanglement-based version of the six-state protocol in the setup configuration presented in Fig.\,\ref{fig:CharlieSource}, $K^\mathrm{CV}/K^\mathrm{6state}$, plotted as a function of the channel transmittance $T$ and the channel noise $\mu$ with the assumption that the utilized sources of light and detection systems are all perfect. (b) Minimal value of the probability $q$ for the realistic source to generate a pair of photons, required for the six-state protocol to produce higher key generation rate than the CV QKD protocol in the ideal case, plotted as a function of $T$ and $\mu$, with the assumption that the photon collection efficiency $\xi=100\%$ and the trusted parties utilize perfect PNR detectors. The key generation rate for the CV QKD [six-state] protocol was calculated with the use of Eq.\,(\ref{eq:KCVsource}) [Eq.\,(\ref{eq:K6state})].}
\label{fig:RatioKeyComparison1}
\end{figure}

Although high resistance to the channel noise is crucial, especially in noisy environments, the main factor determining usefulness of a given QKD protocol for particular situation is the key generation rate it is able to produce. The comparison between these rates that can be obtained in the source-in-the middle setup configuration by utilizing the CV and six-state DV protocols with ideal sources and detectors, plotted for different values of the channel transmittance $T$ and the noise parameter $\mu$, can be seen in Fig.\,\ref{fig:RatioKeyComparison1}\,(a). There are four distinct regions in this plot. In the white region both protocols are insecure. In the blue region the CV protocol is insecure, while the six-state scheme is able to generate secure key at non-zero rate. In the lightest yellow region both protocols are secure and the six-state protocol produces higher key generation rate than the CV protocol. Finally, all the other colors from the intense yellow to red denote the region when both protocols are secure and the six-state protocol produces lower key rate than the CV scheme. The plot in Fig.\,\ref{fig:RatioKeyComparison1}\,(b) is strongly related to these results as it shows the minimal required probability $q$ for the imperfect source to produce a pair of photons in order for the six-state protocol to generate higher key rate than the ideal CV protocol. Here, in the lightest yellow region only the six-state protocol can be secure, so it produces higher key rate for any $q>0$. On the contrary, in the brown region CV protocol produces higher key rate than the six-state protocol even in the case of ideal photon-pair source, so outperforming CV QKD by the six-state protocol is impossible for any $q$. Finally, in the narrow belt colored by the set of colors from intense yellow to red, six-state protocol can beat CV protocol in terms of the key generation rate, but only if $q$ is high enough. Fig.\,\ref{fig:RatioKeyComparison1} was created with the assumption that limited value of $q$ is the only imperfection of the analyzed DV QKD setup. In case of other imperfections these two plots would obviously change in favor of the CV protocol, with the region where $K^\mathrm{CV}/K^\mathrm{6state}>1$ becoming larger and the level of $q$ required for the six-state protocol to generate the key faster than the squeezed-state protocol taking higher values. On the other hand introducing imperfections to the CV setup, or assuming limited error correction efficiency, which for CV case is typically much smaller than for DV QKD, would have the opposite effect. However, the general shape of the two plots would stay similar to what can be seen in Fig.\,\ref{fig:RatioKeyComparison1}. To make this investigation more complete we also compare the performance of all the QKD setup configurations presented in the manuscript in a single figure illustrating the key rate as a function of channel transmittance only, with the assumption that the noise parameter is fixed on either $\mu=0.1$ or $\mu=0.5$ value. Such a comparison in presented in Fig.\,\ref{fig:AdditionalFigure} (b) in Appendix 4.
	
\begin{figure}[tbh]
    \centering
    \includegraphics[width=0.6\textwidth]{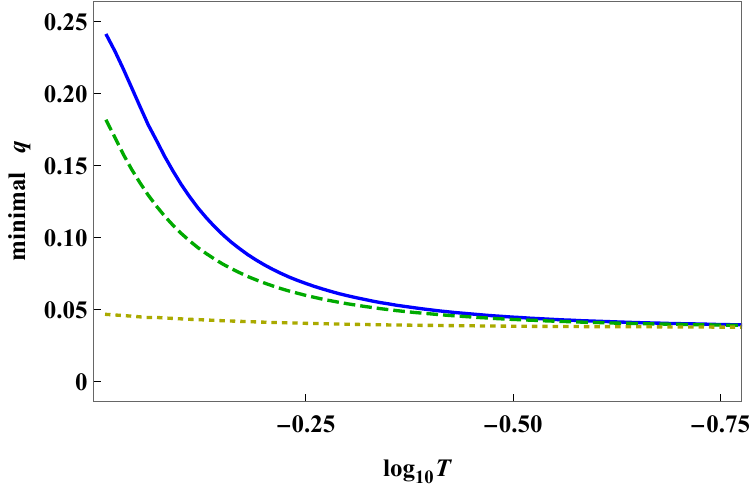} 
    \caption{Minimal value of the probability $q$ for generating a single pair of photons required for the source-in-the-middle scheme, pictured in Fig.\,\ref{fig:CharlieSource}, to exhibit better resistance to the channel noise than the detectors-in-the-middle scheme, shown in Fig.\,\ref{fig:CharlieDetectors}, when Alice and Bob utilize analogs of the six-state protocol, plotted as a function of the channels' transmittance $T$. The blue solid line (green dashed line) is obtained for the case of ideal PNR detectors utilized by the trusted parties and the photon collection efficiency of the sources equal to $\xi=100\%$ ($\xi=80\%$). The yellow dotted line corresponds to the case of the ideal on/off detectors and $\xi=100\%$.}
\label{fig:DVSchemesComparisonRealistic}
\end{figure}

The considered deterministic type of photon pair sources, combined with the noiseless detection performed by Alice and Bob, makes the detectors-in-the middle scheme pictured in Fig.\,\ref{fig:CharlieDetectors} totally independent of $q$ in terms of the maximal tolerable channel noise. This is because in this situation the local detectors placed in the trusted parties' laboratories can click only if single pairs were generated by their respective sources, so all the empty-pulse events are automatically discarded from the key generation process. On the other hand, the resistance of the source-in-the-middle scheme to the channel noise deteriorates with decreasing $q$, as shown in the previous paragraphs. Thus, the following question is justified: what is the minimal value of the parameter $q$ needed for the source-in-the-middle setup configuration to be more resistant to the channel noise than the detectors-in-the middle one in the DV case? The answer to this question largely depends on the ability of the utilized single-photon sources to resolve the number of detected photons, as can be seen in Fig.\,\ref{fig:DVSchemesComparisonRealistic}. When the detectors are PNR and the channel transmittance is very high, the required value of the single pair generation probability reaches the highest level of about $24.5\%$. Then, with decreasing $T$ it asymptotically drops to slightly below $4\%$. However, when the detectors are of the on/off type, this threshold value of $q$ stays below $5\%$ even for the highest $T$. In the PNR case non-ideal collection efficiency of the generated photons also reduces the required photon-pair generation efficiency in significant way. Taking all of it into account it can be concluded that in realistic situations one should expect the source-in-the-middle setup configuration to perform better than the detectors-in-the middle one unless the photon-pair sources are of really poor quality.
	
\section{Summary}
\label{Sec:Summary}

In this work we analyzed and compared the performance of the entanglement-based versions of various DV and CV QKD protocols, implemented with different setup configurations, in the conditions of lossy and noisy quantum channels, with otherwise perfect setups. We derived the fundamental bounds of the noise and transmittance of quantum channels for entanglement-based CV QKD protocols and found out that both in the source-in-the-middle and detectors-in-the-middle configurations the analogs of the six-state and BB84 protocols offer better resistance to the channel noise than the CV protocol, regardless of the transmittance of the utilized channels. Moreover, due to the fundamental 3 dB limit on the loss on the reference side in the CV QKD \cite{Usenko2016}, the CV protocol becomes insecure for relatively high transmittance even in the absence of channel noise. This is in contrast to the analyzed DV protocols, which can remain secure even for very small $T$ as long as the channel noise is not too strong. 
	
Motivated by these results we investigated the influence of various setup imperfections on the DV QKD security in order to determine the benchmarks for several parameters describing the quality of realistic photon-pair sources and detectors, guaranteeing the superiority of the six-state protocol over even the ideal CV protocol implementations. The benchmarks turned out to be relatively demanding for the currently existing photon-pair sources, especially in terms of the required collection efficiency of the generated photons. For high-transmittance channel this efficiency needs to exceed $60\%$ even in the case when all of the other setup parameters are nearly perfect. Moreover, for high-transmittance links outperforming the ideal CV protocol by the realistic six-state protocol may become impossible when the trusted parties are equipped with binary on/off detectors, without photon-number-resolution. Nevertheless, for the channel transmittance of the order of $65\%-75\%$ realistic setup components that are already accessible nowadays may be enough for the DV protocols to provide their users with higher resistance to the channel noise than the ideal CV QKD. The ongoing technological progress in the fields of photon-pair generation and detection may also allow it for higher transmittance values in the near future.
	
To make our investigation complete, we also compared the key generation rates that can be obtained by the six-state and CV protocols for the quantum channels with different levels of loss and noise. We identified the regions on the plot where one of these two protocols performs strictly better than the other one and the area where the DV scheme can provide the trusted parties with larger key rate than the CV scheme as long as the utilized photon-pair source is of sufficiently high quality. Finally, we investigated the possibility for the detectors-in-the-middle DV QKD configuration to offer better resistance to the channel noise than the analogous source-in-the-middle scheme, discovering that it can be realized only in the case of using very inefficient sources in their implementations.
	
Our work provides valuable insight to the realistic performance and its theoretical limits for various QKD protocols implemented in the entanglement-based setup configurations. In particular, we focused our attention on the rarely explored regime of high channel transmittance and strong noise. Such regime can nevertheless be relevant in practical situations, such as co-existence with strong classical signals, where crosstalk effects result in excess noise \cite{Eraerds2010}. Therefore, the presented results can become very useful in estimating the expected performance and choosing the optimal solutions for the short QKD links realized in very noisy environment. Since the theoretical analysis can be easily adapted to cover various types of channel noise, it can be also used as a base for the evaluation of local quantum-classical communication networks populated with high traffic \cite{Frohlich2013,Tang2016,Sun2018,Dynes2019}.
	
\section*{Acknowledgements}

The authors acknowledge fruitful discussions with Prof. Radim Filip and support from the project ``Secure quantum communication in multiplexed optical networks'' run in co-operation by the National Science Centre (NCN) in Poland and the Czech Science Foundation (GA\v{C}R). M.L.\,was supported by the grant no.\,2020/39/I/ST2/02922 of NCN, O.K.\,and V.C.U.\,were supported by the grant no.\,21-44815L of GA\v{C}R. O.K. and V.C.U. acknowledge the project 8C22003 (QD-E-QKD) of MEYS of Czech Republic, which has received funding from the European
Union’s Horizon 2020 research and innovation framework programme under Grant Agreement No. 731473 and 101017733; EU H2020-WIDESPREAD-2020-5 project
NONGAUSS (951737) under the CSA-Coordination and Support Action; "Quantum Security Networks Partnership" (QSNP, grant agreement No. 101114043) project
funded from the European Union’s Horizon Europe research and innovation programme.   
		
\section*{Appendix 1}
\label{Sec:App1}
	
Assuming that efficiency of the detectors placed at the central station is given by $\eta_C$ the joint probability for detecting $Y$-polarized photon by Alice, $Z$-polarized photon by Bob and receiving a pair of clicks in the detectors $\mathrm{D}_{1\mathrm{H}}$ and $\mathrm{D}_{1\mathrm{V}}$ takes the following form:
\begin{align}
    &P_\mathrm{YZ}^\mathrm{1H,1V}=\frac{1}{4}\sum_{i=0}^\infty\sum_{j=0}^\infty\sum_{k=0}^\infty\sum_{l=0}^\infty\frac{\mu_A^{i+j}\mu_B^{k+l}\eta_C^{i+j+k+l}}{i!\,j!\,k!\,l!\,(\mu_A\eta_C+1)^{i+j+2}(\mu_B\eta_C+1)^{k+l+2}}\sum_{\alpha=0}^1\sum_{\beta=0}^1(\xi_A\eta_C)^\alpha(\xi_B\eta_C)^\beta\nonumber\\&\times(1-\xi_A\eta_C)^{1-\alpha}(1-\xi_B\eta_C)^{1-\beta}\sum_{\gamma=0}^\alpha\sum_{\delta=0}^\beta\sum_{\sigma=0}^\alpha\sum_{\tau=0}^\beta(-1)^{\gamma+\delta-\sigma-\tau}f_\mathrm{YZ},
\end{align}
where $f_\mathrm{YZ}$ depends on the type of detectors utilized in the central station's measurement system. In the PNR case
\begin{equation}
    f_\mathrm{HH}^\mathrm{PNR}=\frac{1}{4}\sum_{i'=\mathrm{Max}[0,1-k]}^{\mathrm{Min}[i,1]}\sum_{j'=\mathrm{Max}[0,\sigma-\gamma,1-l-\gamma-\delta,1-l-\gamma-\tau]}^{\mathrm{Min}[j,j+\sigma-\gamma,1-\gamma-\delta,1-\gamma-\tau]}C\,g^\mathrm{PNR}h_\mathrm{HH}^\mathrm{PNR}
\end{equation}
and
\begin{equation}
    f_\mathrm{HV}^\mathrm{PNR}=\frac{1}{4}\sum_{i'=\mathrm{Max}[0,1-k-\delta,1-k-\tau]}^{\mathrm{Min}[i,1-\delta,1-\tau]}\sum_{j'=\mathrm{Max}[0,\sigma-\gamma,1-l-\gamma]}^{\mathrm{Min}[j,j+\sigma-\gamma,1-\gamma]}C\,g^\mathrm{PNR}h_\mathrm{HV}^\mathrm{PNR},
\end{equation}
where the newly introduced functions are as follows:
\begin{equation}
    C={i \choose i'}^2{j \choose j'}{j \choose j'+\gamma-\sigma}(i-i')!\,(j-j'+\alpha-\gamma)!\,t_1^{i+j-i'-j'+\sigma}(1-t_1)^{\alpha+i'+j'-\sigma},
\end{equation}
\begin{equation}
    g^\mathrm{PNR}=t_2^{k+l+\gamma+\delta+\tau+i'+j'-2}(1-t_2)^{\beta-\gamma-\delta-\tau-i'-j'+2},
\end{equation}
\begin{equation}
    h_\mathrm{HH}^\mathrm{PNR}={k \choose 1-i'}^2{l \choose 1-j'-\gamma-\delta}{l \choose 1-j'-\gamma-\tau}(k+i'-1)!\,(l+j'+\gamma-1+\beta)!,
\end{equation}
\begin{equation}
    h_\mathrm{HV}^\mathrm{PNR}={k \choose 1-i'-\delta}{k \choose 1-i'-\tau}{l \choose 1-j'-\gamma}^2(k+i'-1+\beta)!\,(l+j'+\gamma-1)!.
\end{equation}
	
On the other hand in the case of binary on/off detectors:
\begin{equation}
    f_\mathrm{HH}^\mathrm{on/off}=\sum_{i'=0}^{i}\sum_{j'=\mathrm{Max}[0,\sigma-\gamma]}^{\mathrm{Min}[j,j+\sigma-\gamma]}\sum_{k'=\mathrm{Max}[0,1-i']}^{k}\sum_{l'=\mathrm{Max}[0,\tau-\delta,1-\gamma-\delta-j']}^{\mathrm{Min}[l,l+\tau-\delta]}C\,g^\mathrm{on/off}h_\mathrm{HH}^\mathrm{on/off}
\end{equation}
and
\begin{equation}
	f_\mathrm{HV}^\mathrm{on/off}=\sum_{i'=0}^{i}\sum_{j'=\mathrm{Max}[0,\sigma-\gamma]}^{\mathrm{Min}[j,j+\sigma-\gamma]}\sum_{k'=\mathrm{Max}[0,\tau-\delta,1-\delta-i']}^{\mathrm{Min}[k,k+\tau-\delta]}\sum_{l'=\mathrm{Max}[0,1-\gamma-j']}^lC\,g^\mathrm{on/off}h_\mathrm{HV}^\mathrm{on/off},
\end{equation}
where
\begin{equation}
    g^\mathrm{on/off}=\frac{t_2^{k+l-k'-l'+\tau}(1-t_2)^{\beta+k'+l'-\tau}}{2^{\gamma+\delta+i'+j'+k'+l'}},
\end{equation}
\begin{equation}
    h_\mathrm{HH}^\mathrm{on/off}={k \choose k'}^2{l \choose l'}{l \choose l'+\delta-\tau}(k-k')!\,(l-l'+\beta-\delta)!\,(i'+k')!\,(j'+l'+\gamma+\delta)!
\end{equation}
and
\begin{equation}
    h_\mathrm{HV}^\mathrm{on/off}={k \choose k'}{k \choose k'+\delta-\tau}{l \choose l'}^2(k-k'+\beta-\delta)!\,(l-l')!\,(i'+k'+\delta)!\,(j'+l'+\gamma)!.
\end{equation}
	
\section*{Appendix 2}

\begin{figure}[tbh]
    \centering
    \includegraphics[width=1.0\textwidth]{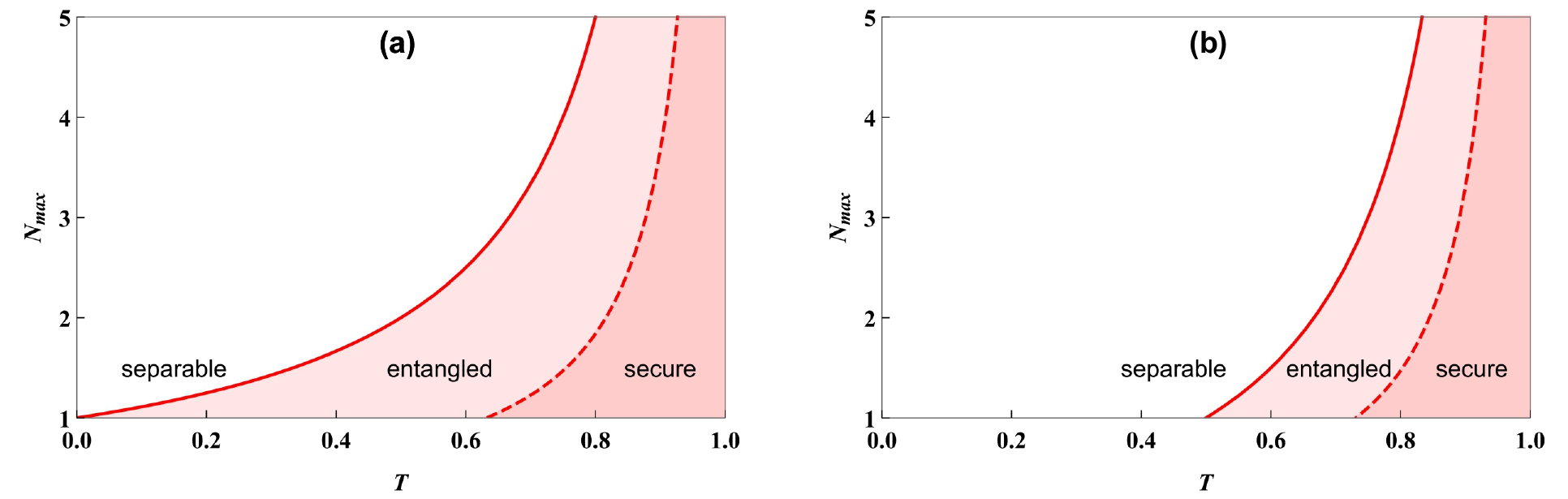} 
    \caption{Maximum tolerable channel noise $N_{max}$ for the Gaussian logarithmic negativity (red solid lines) and for the key rate (red dashed lines), plotted as functions of the channel attenuation $T$ for standard EPR-based scheme (left panel) and for the MDI scheme (right panel). All the plots are made with the assumption that the utilized source(s) of entanglement are ideal, $V\rightarrow\infty$. Noise levels above the solid lines make the state separable by breaking the entanglement. Below the solid lines the state remains entangled. Below the dashed lines the noise level is compatible with secure QKD.}
\label{fig:LNvsKR}
\end{figure}
	
Here we describe the evaluation of logarithmic negativity for CV schemes with source of entanglement in the middle (standard EPR-based scheme) and with detectors in the middle of the channel (MDI scheme). Logarithmic negativity (LN) \cite{Vidal2002} is an entanglement monotone, which indicates to what extent the partially transposed state fails to be physical. Gaussian LN can be easily calculated through the minimum symplectic eigenvalue, $\nu_-$, of the partially transposed covariance matrix as $LN=max\{0,-\log_2 \nu_-\}$ \cite{Serafini2004} (with no loss of generality we take logarithm to the base 2, similarly to entropic quantities used for evaluation of the key rate). 
	
In the source-in-the-middle case straightforward evaluation of LN from the partial transpose of covariance matrix (\ref{eq:covmat_source}), assuming the same transmittance and noise in both arms of the entangled state, $T_A = T_B \equiv T$, $N_A = N_B \equiv N$, gives
\begin{equation}
    LN=-\frac{1}{2}\log_2\big[
    N^2(1-T)^2+2N(1-T)TV-2T[TV+N(1-T)]\sqrt{V^2-1}+T^2(2V^2-1)
    \big].
\end{equation}
Entanglement is broken when the expression given above turns to zero. It happens at the maximum thermal noise level of
\begin{equation}
    N_{max}=\frac{T\sqrt{V^2-1}-TV+1}{1-T},
\end{equation}
which in the limit of infinitely strong entanglement $V \to \infty$ reduces to $N_{max}=1/(1-T)$. Similarly, for the purely attenuating channel, $N=1$, in the limit of $V \to \infty$ entanglement turns to $LN=-\log_2{(1-T)}$. 
	
For the MDI scheme the expression for the LN, evaluated from the partial transpose of the matrix (\ref{eq:covmat_det}) in the symmetric channel is 
\begin{equation}
    LN=\log_2{\frac{TV+N(1-T)}{T+NV(1-T)}}.
\end{equation}
In the limit of $V \to \infty$ it reads $LN=\log_2{T}-\log_2{[N(1-T)]}$. If the channel noise is present, the entanglement for the MDI scheme is broken by
\begin{equation}
    N_{max}=\frac{T}{1-T},
\end{equation}
which is independent on the state variance $V$. 
	
The bounds on the channel noise limiting the Gaussian entanglement of the shared states in either of the configurations, found above, are plotted in Fig. \ref{fig:LNvsKR} for the case when the utilized sources of entanglement are ideal. Similar bounds for breaking the QKD security of these schemes (turning the key rate to zero) are also illustrated there for comparison. It is evident from the plots, that entanglement in the MDI scheme is more sensitive to the channel loss and so is the key rate, compared to the standard EPR-based protocol. 
	
\section*{Appendix 3}
\label{Sec:App2}

\begin{figure}[tbh]
    \centering
    \includegraphics[width=0.6\textwidth]{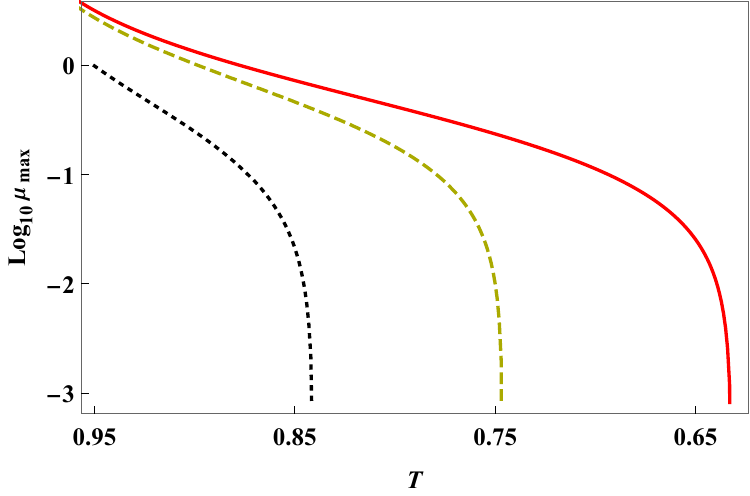} 
    \caption{Maximal values of the channel noise $\mu$ for which it is possible to generate the secure key, plotted as a function of channel transmittance $T$ calculated numerically for the cases of Alice and Bob utilizing DI QKD protocol presented in Ref.\,\cite{Acin07} in the setup configuration shown in Fig.\,\ref{fig:CharlieSource} (black dotted line), CV protocol in the setup configuration shown in Fig.\,\ref{fig:CharlieSourceCV} (red solid line) and CV protocol in the setup configuration shown in Fig.\,\ref{fig:CharlieDetectorsCV} (yellow dashed line). The plots for CV protocols were made with the assumption that the sources and detection systems are ideal.}
\label{fig:DIwithCVSchemesComparison}
\end{figure}

The setup configuration presented in Fig.\,\ref{fig:CharlieSource} can be also utilized for QKD in the device-independent (DI) scenario, where it is assumed that not only the source placed in the central station, but also the detection schemes used by Alice and Bob are all untrusted \cite{Acin07}. In this case the legitimate parties check the security of their key generation process by attempting to violate the CHSH inequality \cite{Clauser69}. It can be done for appropriate set of rotation angles $\{\theta_1^A,\theta_2^A\}$ and $\{\theta_1^B,\theta_2^B\}$ chosen for their respective polarization modulators. Additionally, one of the parties, say Bob, should also use the third angle $\theta_0^B=\theta_1^A$. During the process of key generation the legitimate parties randomly switch between the angles available for their modulators. Then, in the stage of the basis reconciliation the results obtained by them when Alice's angle was $\theta_1^A$ ($\theta_2^A$) and Bob's angle was $\theta_0^B$ are used to form the key (discarded), while all of the other measurement results are used for the CHSH inequality check. In the asymptotic case of infinite key the probability for the legitimate parties to use the combination of angles $(\theta_1^A,\theta_0^B)$ can be made arbitrarily close to one.
	
In this work we calculated the expected performance of the DI QKD protocol in the case when the only imperfect setup elements are the quantum channels connecting the central station with Alice and Bob. With our standard model for the channel noise, described in \cite{Lasota2017}, the joint probability for registering $n_1$, $n_2$, $n_3$ and $n_4$ photons by the detectors $\mathrm{D}_\mathrm{AH}$, $\mathrm{D}_\mathrm{AV}$, $\mathrm{D}_\mathrm{BH}$ and $\mathrm{D}_\mathrm{BV}$, respectively, in a single event, conditioned on the combination of angles $(\theta^A,\theta^B)$, chosen by Alice and Bob, can be written as 
\begin{align}
    &P_{n_1,n_2,n_3,n_4}(\theta^A,\theta^B)=\frac{n_1!\,n_2!\,n_3!\,n_4!}{2}\sum_{i=0}^1\sum_{j=0}^1(-1)^{i+j}\sum_{m_1=0}^\infty\sum_{m_2=0}^\infty\sum_{m_3=0}^\infty\sum_{m_4=0}^\infty\frac{1}{m_1!\,m_2!\,m_3!\,m_4!}\,\,\,\,\,\,\,\,\,\,\,\,\,\,\nonumber\\&\times\frac{\mu_1^{m_1+m_2}\mu_2^{m_3+m_4}}{(\mu_1+1)^{m_1+m_2+2}(\mu_2+1)^{m_3+m_4+2}}\sum_{m_1'=0}^{m_1}\sum_{m_2'=0}^{m_2}\sum_{m_3'=0}^{m_3}\sum_{m_4'=0}^{m_4}{m_1 \choose m_1'}{m_2 \choose m_2'}{m_3 \choose m_3'}{m_4 \choose m_4'}\nonumber\\&\times\sum_{i'=\mathrm{Max}[0,n_1+n_2+i-m_1-m_2+m_1'+m_2'-1]}^{\mathrm{Min}[i,n_1+n_2-m_1-m_2+m_1'+m_2']}\,\sum_{i''=\mathrm{Max}[0,n_3+n_4-i-m_3-m_4+m_3'+m_4']}^{\mathrm{Min}[1-i,n_3+n_4-m_3-m_4+m_3'+m_4']}(i+m_1'-i')!\nonumber\\&\times(1-n_1-n_2-i+m_1+m_2-m_1'+i')!\,(1-i+m_3'-i'')!\,(i-n_3-n_4+m_3+m_4-m_3'+i'')!\nonumber\\&\times\sum_{m_1''=\mathrm{Max}[0,i-j+m_1'-i']}^{\mathrm{Min}[m_1,i+m_1'-i']}\,\sum_{m_2''=\mathrm{Max}[0,-n_1-n_2-i+j+m_1+m_2-m_1'+i']}^{\mathrm{Min}[m_2,1-n_1-n_2-i+m_1+m_2-m_1'+i']}\,\sum_{m_3''=\mathrm{Max}[0,j-i+m_3'-i'']}^{\mathrm{Min}[m_3,1-i+m_3'-i'']}\nonumber\\&\times\sum_{m_4''=\mathrm{Max}[0,-n_3-n_4+i-j+m_3+m_4-m_3'+i'']}^{\mathrm{Min}[m_4,-n_3-n_4+i+m_3+m_4-m_3'+i'']}v_1(n_1,n_2,n_3,n_4,i,j,m_1,m_3,m_1',m_3',m_1'',m_3'',i',i'',\theta^A,\theta^B)\nonumber\\&\times v_2(n_1,n_2,n_3,n_4,m_1,m_2,m_3,m_4,m_1',m_2',m_3',m_4',m_1'',m_2'',m_3'',m_4''),
\end{align}
where
\begin{align}
    &v_1(n_1,n_2,n_3,n_4,i,j,m_1,m_3,m_1',m_3',m_1'',m_3'',i',i'',\theta^A,\theta^B)=\sum_{\alpha=\mathrm{Max}[0,-n_2+m_1-m_1'+i']}^{\mathrm{Min}[n_1,m_1-m_1'+i']}\nonumber\\&\times\sum_{\beta=\mathrm{Max}[0,-n_2-i+j+m_1-m_1'+i']}^{\mathrm{Min}[n_1,-i+j+m_1-m_1'+i']}\,\sum_{\gamma=\mathrm{Max}[0,-n_4+m_3-m_3'+i'']}^{\mathrm{Min}[n_3,m_3-m_3'+i'']}\,\sum_{\delta=\mathrm{Max}[0,-n_4+i-j+m_3-m_3'+i'']}^{\mathrm{Min}[n_3,i-j+m_3-m_3'+i'']}(-1)^{\alpha+\beta+\gamma+\delta}\nonumber\\&\times{m_1-m_1'+i' \choose \alpha}{-i+j+m_1-m_1'+i' \choose \beta}{m_3-m_3'+i'' \choose \gamma}{i-j+m_3-m_3'+i'' \choose \delta}\nonumber\\&\times{n_1+n_2-m_1+m_1'-i' \choose n_1-\alpha}{n_1+n_2+i-j-m_1+m_1'-i' \choose n_1-\beta}{n_3+n_4-m_3+m_3'-i'' \choose n_3-\gamma}\nonumber\\&\times{n_3+n_4-i+j-m_3+m_3'-i'' \choose n_3-\delta}\left(\sin\theta^A\right)^{-i+j+2(n_1+m_1-m_1'+i'-\alpha-\beta)}\nonumber\\&\times\left(\cos\theta^A\right)^{i-j+2(n_2-m_1+m_1'-i'+\alpha+\beta)}\left(\sin\theta^B\right)^{i-j+2(n_3+m_3-m_3'+i''-\gamma-\delta)}\nonumber\\&\times\left(\cos\theta^B\right)^{-i+j+2(n_4-m_3+m_3'-i''+\gamma+\delta)}
\end{align}
and
\begin{align}
    &v_2(n_1,n_2,n_3,n_4,m_1,m_2,m_3,m_4,m_1',m_2',m_3',m_4',m_1'',m_2'',m_3'',m_4'')={m_1 \choose m_1''}{m_2 \choose m_2''}{m_3 \choose m_3''}{m_4 \choose m_4''}\nonumber\\&\times(-1)^{m_1'+m_2'+m_3'+m_4'-m_1''-m_2''-m_3''-m_4''}\,\,T_A^{\,n_1+n_2-m_1-m_2+m_1'+m_2'+m_1''+m_2''}\nonumber\\&\times\left(1-T_A\right)^{1-n_1-n_2+2m_1+2m_2-m_1'-m_2'-m_1''-m_2''}\,T_B^{\,n_3+n_4-m_3-m_4+m_3'+m_4'+m_3''+m_4''}\nonumber\\&\times\left(1-T_B\right)^{1-n_3-n_4+2m_3+2m_4-m_3'-m_4'-m_3''-m_4''}.
\end{align}
Since errors in the raw key are generated when clicks are registered only in the pairs of detectors $\mathrm{D}_\mathrm{AH}$ and $\mathrm{D}_\mathrm{BH}$ or $\mathrm{D}_\mathrm{AV}$ and $\mathrm{D}_\mathrm{BV}$, QBER can be calculated as
\begin{equation}
    Q=\frac{\sum_{x=1}^\infty\sum_{y=1}^\infty \left[P_{x,0,y,0}(\theta^A_1,\theta^B_0)+P_{0,x,0,y}(\theta^A_1,\theta^B_0)\right]}{p_\mathrm{exp}},
\end{equation}
where
\begin{equation}
    p_\mathrm{exp}=\sum_{x=1}^\infty\sum_{y=1}^\infty \left[P_{x,0,y,0}(\theta^A_1,\theta^B_0)+P_{0,x,0,y}(\theta^A_1,\theta^B_0)+P_{x,0,0,y}(\theta^A_1,\theta^B_0)+P_{0,x,y,0}(\theta^A_1,\theta^B_0)\right].
\end{equation}
Furthermore, the CHSH combination is defined as
\begin{equation}
	S=E(\theta_1^A,\theta_1^B)+E(\theta_1^A,\theta_2^B)+E(\theta_2^A,\theta_1^B)-E(\theta_2^A,\theta_2^B),
\end{equation} 
where
\begin{equation}
    E(\theta^A,\theta^B)=\sum_{x=1}^\infty\sum_{y=1}^\infty \left[P_{x,0,y,0}(\theta^A,\theta^B)+P_{0,x,0,y}(\theta^A,\theta^B)-P_{x,0,0,y}(\theta^A,\theta^B)-P_{0,x,y,0}(\theta^A,\theta^B)\right].
\end{equation}
When the values of $Q$, $p_\mathrm{exp}$ and $S$ are known, they can be used to calculate the lower bound for the achievable key generation rate in the case of Alice and Bob utilizing DI QKD protocol described above. This bound is given by the expression \cite{Acin07}:
\begin{equation}
    K^\mathrm{(DI)}=p_\mathrm{exp}\max\!\!\left[0,1-H(Q)-H\left(\frac{1+\sqrt{(S/2)^2-1}}{2}\right)\right].
\end{equation}
	
The comparison between the DI QKD protocol and the CV protocol realized with the setups illustrated in Fig.\,\ref{fig:CharlieSourceCV} and Fig.\,\ref{fig:CharlieDetectorsCV} in terms of their maximal tolerable channel noise is presented in Fig.\,\ref{fig:DIwithCVSchemesComparison}. As could be expected the DI QKD scheme, offering the highest level of security, places the most demanding limitation on $\mu$, regardless of the transmittance of the quantum channels. Also the lowest $T$ for which the security of the DI QKD protocol is still possible is significantly higher than in the case of the CV schemes, approaching the value of $0.84$. The main reason for this is the stringent requirement on the efficiency of the setup that has to be fulfilled in order for the violation of the CHSH inequality to be possible \cite{Mermin86,Brunner2014}.
	
\section*{Appendix 4}

\begin{figure}[tbh]
    \centering
    \includegraphics[width=1.0\textwidth]{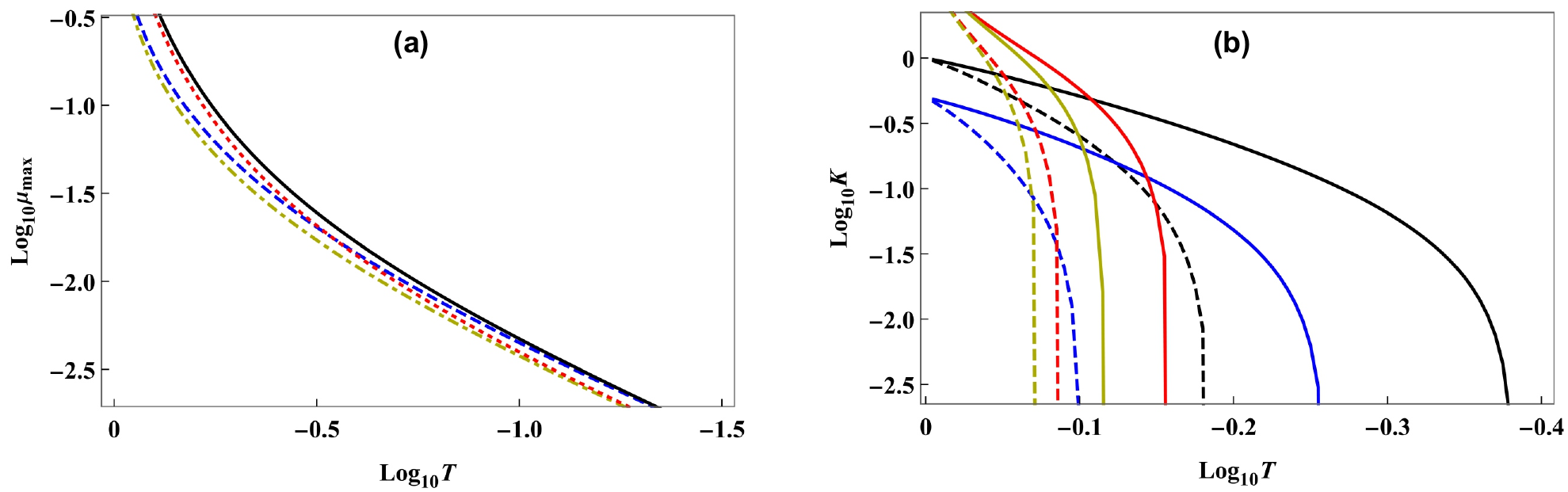} 
    \caption{(a) Maximal values of the channel noise $\mu$ for which it is possible to generate the secure key, plotted as a function of channel transmittance $T$, calculated numerically for the cases of Alice and Bob utilizing entanglement-based version of the six-state [BB84] protocol in the setup configuration shown in Fig.\,\ref{fig:CharlieDetectors}, with PNR detectors (black solid [red dotted] line) or binary on/off detectors (blue dashed [yellow dot-dashed] line). All the plots were made with the assumptions that the detection efficiency is $\eta=100\%$ and the photon-pair sources utilized by Alice and Bob are ideal. (b) Lower bound for the key generation rate that can be obtained by realizing the six-state protocol with the schemes illustrated in Fig.\,\ref{fig:CharlieSource} (black lines) and Fig.\,\ref{fig:CharlieDetectors} (blue lines) or the squeezed-state protocol with the schemes illustrated in Fig.\,\ref{fig:CharlieSourceCV} (red lines) and Fig.\,\ref{fig:CharlieDetectorsCV} (yellow lines), plotted as a function of channel transmittance. Solid (dashed) lines indicate the results obtained for $\mu=0.1$ ($\mu=0.5$). For all the plots the sources and detectors are assumed to be perfect.}
\label{fig:AdditionalFigure}
\end{figure}

In this appendix we present two additional figures that can make the presented analysis more complete, although we consider them not necessary for the discussion of the main results of our work. In Fig.\,\ref{fig:AdditionalFigure} (a) one can see an analogous comparison between different DV QKD protocols realized with either PNR or binary on/off detectors to the one illustrated in Fig.\,\ref{fig:MainPlotsIdealSources} (b), but calculated for the detectors-in-the-middle setup configuration shown in Fig.\,\ref{fig:CharlieDetectors}. The obtained results are very similar to Fig.\,\ref{fig:MainPlotsIdealSources} (b) and the conclusions that can be drawn are the same. On the other hand, in Fig.\,\ref{fig:AdditionalFigure} (b) we present a standard key rate vs. channel transmittance type of plot, where we compare the performance of all the setup configurations shown in Figs.\,\ref{fig:CharlieSource}--\ref{fig:CharlieDetectorsCV}, utilized for the realization of either six-state or squeezed-state protocol, assuming two different values of the noise parameter $\mu$. The figure confirms the general superiority of the source-in-the-middle type of QKD schemes, discussed in the main text. It is also compatible with Fig.\,\ref{fig:RatioKeyComparison1}, showing that while for near-unity transmittance of quantum channels CV protocols are capable of providing significantly higher key generation rate than their DV analogs, they become insecure much quicker when $T$ decreases.
	
\bibliography{NJPsubmissionBiblio}

\end{document}